\documentclass[onecolumn,11pt,draftclsnofoot,peerreview]{IEEEtran}

\usepackage{color}
\usepackage{graphics}                       
\usepackage[dvips]{graphicx}
\usepackage[ansinew]{inputenc}              
\usepackage{enumerate}
\usepackage{subfigure}
\usepackage{cite}
\usepackage{amssymb}
\usepackage{amsmath}
\usepackage{amsfonts}
\newtheorem{example}{\bf Example}

\begin{document}
\title{A New Formal Approach for Predicting Period Doubling Bifurcations in Switching Converters}
\author{A. ~El~Aroudi
\thanks{The author is with the Departament d'Enginyeria Electrònica, Elèctrica i Automàtica, Escola Tècnica Superior d'Enginyeria, Universitat Rovira i Virgili, 43007, Tarragona, Spain (e-mails:abdelali.elaroudi@urv.cat).}
}

\date{}

\maketitle

\begin{abstract}
Period doubling bifurcation leading to subharmonic oscillations are undesired phenomena in switching converters. In past studies, their prediction has been mainly
tackled by explicitly deriving a discrete time model and then linearizing it in the vicinity of the operating point. However, the results obtained from such an approach
cannot be applied for design purpose. Alternatively, in this paper, the subharmonic oscillations in voltage mode controlled DC-DC buck converters are predicted by using
a formal symbolic approach. This approach is based on expressing the subharmonic oscillation conditions in the frequency domain and then converting the results to
generalized hypergeometric functions. The obtained expressions depend explicitly on the system parameters and the operating duty cycle making the results directly
applicable for design purpose. Under certain practical conditions concerning these parameters, the hypergeometric functions can be approximated by polylogarithm and
standard functions. The new approach is demonstrated using an example of voltage-mode-controlled buck converters. It is found that the stability of the converter is
strongly dependent upon a polynomial function of the duty cycle.
\end{abstract}


\newpage
\section{Nomenclature}
Let us define the following parameters related to the buck switching regulator studied in this work and that are listed in analphabetic order.
\begin{center}
\begin{tabular}{ll}
$\alpha=\dfrac{\omega_{z2}}{\omega_0}$ & Ratio between $\omega_{z2}$  and ${\omega_0}$\\
$C$     &           Output Capacitance\\
$D$ & Steady state duty cycle\\
$f_s$ & Switching frequency\\
$\omega_{ z1}=\dfrac{1}{r_cC}$ & Zero due to the ESR $r_C$\\
$k_v$ &  Proportional gain\\
$L$     &           Inductance\\
$Q_0=\dfrac{\sqrt{LC(R+r_c)(R+r_\ell)}}{C(Rr_c+Rr_\ell+r_cr_\ell)+L}$ & Damping quality factor\\
$R$     &           Load resistance\\
$r_c$   &           ESR of the capacitance $C$\\
$r_\ell$&           ESR of the inductance  $L$\\
$T$ & Switching period\\
$v_g$&                 Input source voltage\\
$\omega_0=\sqrt{\dfrac{R+r_\ell}{(R+r_c)LC}}$ & Angular resonance frequency\\
$\omega_{1p}$ & Pole of the type II controller\\
$\omega_s=2\pi f_s$ & Angular switching frequency\\
$\omega_{z1}$ & Zero due to the ESR of output the capacitor $C$\\
$\omega_{z2}$ & Zero of the PI controller

\end{tabular}
\end{center}

\section{Introduction}
\begin{figure*}
\begin{center}
\subfigure[]{\includegraphics[width=8cm]{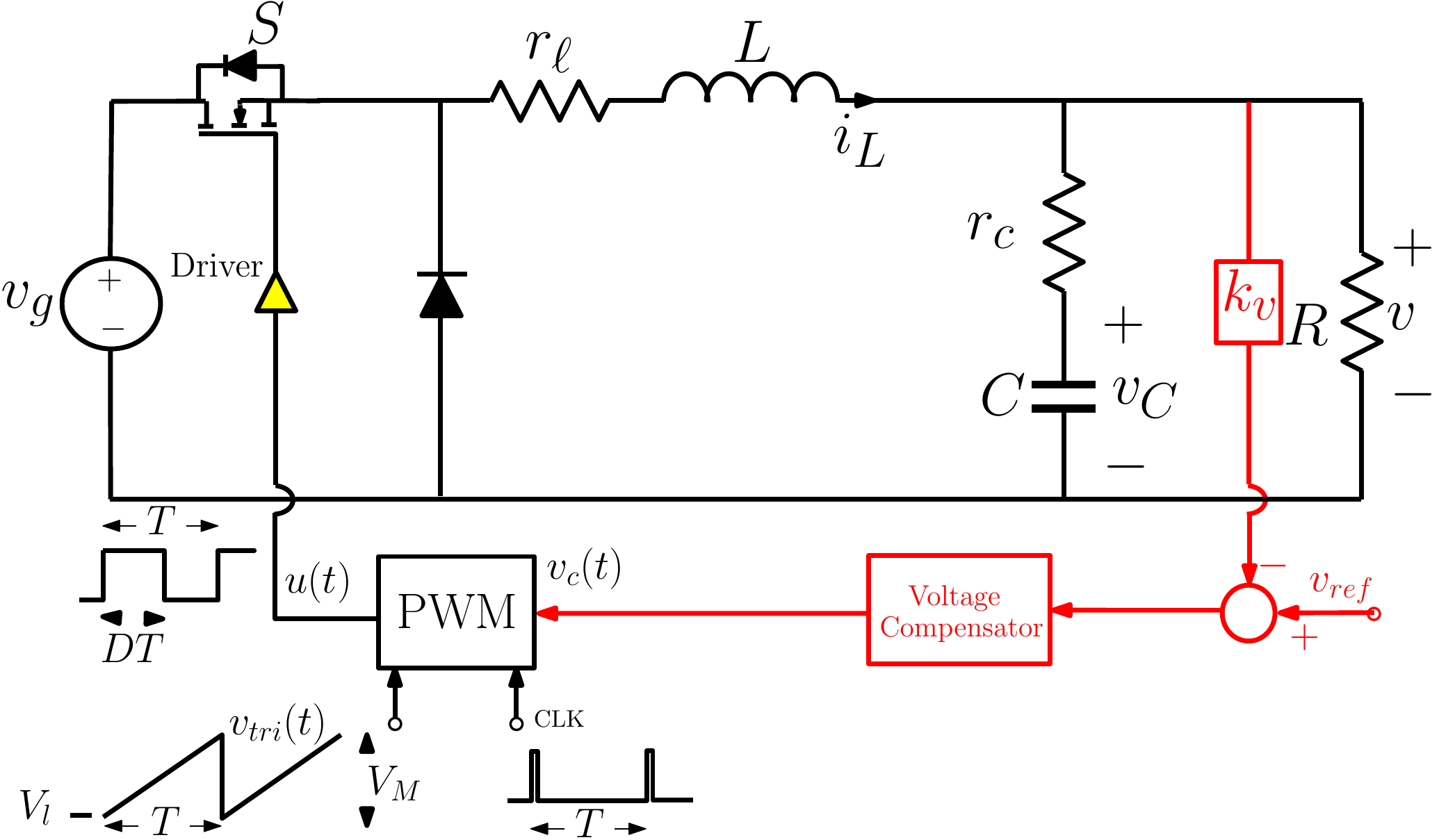}} \subfigure[]{\includegraphics[width=8cm]{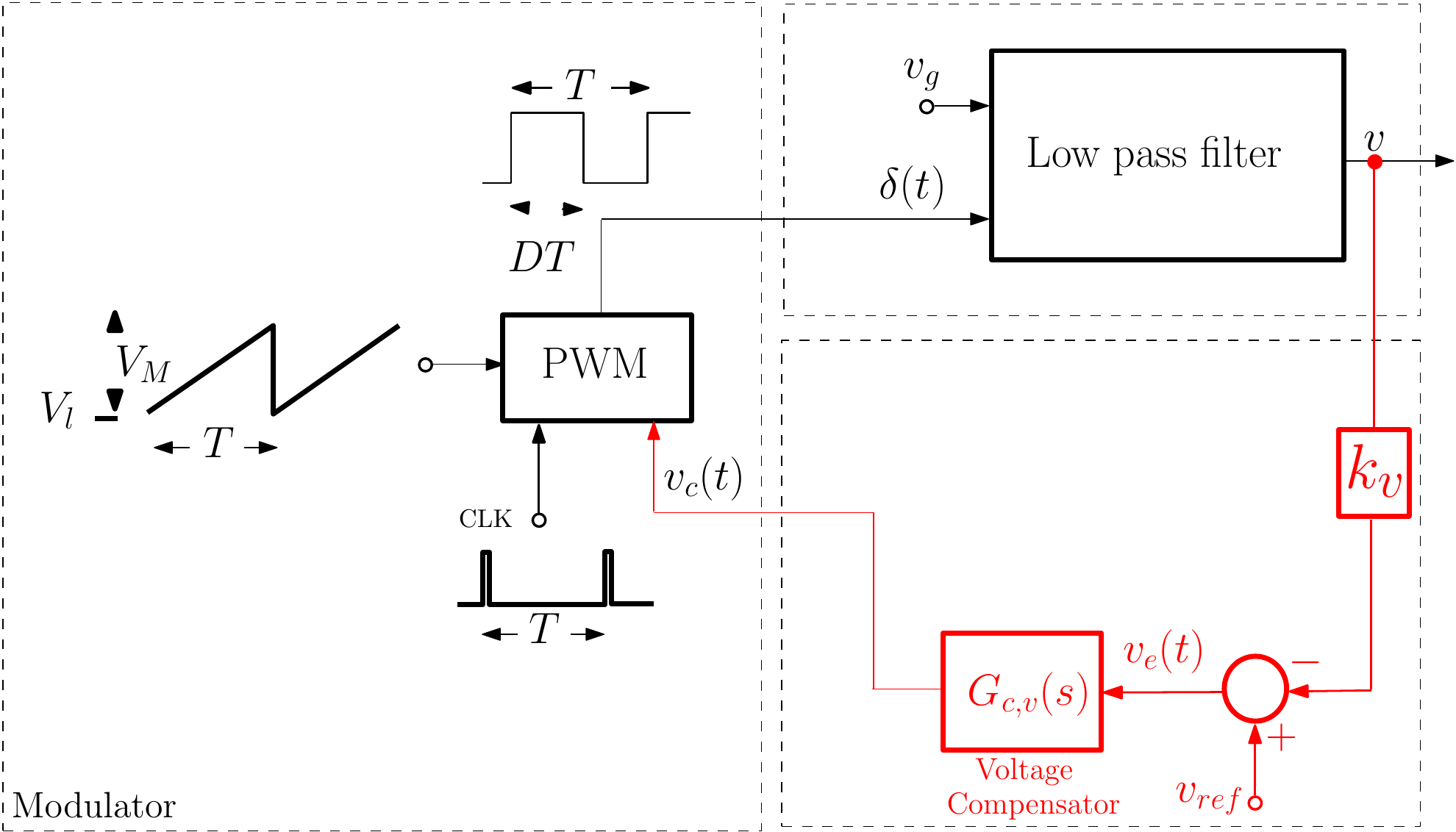}} \caption{(a) Block circuit diagram a DC-DC buck converter under
voltage mode control with a controller with transfer function $G_c(s)$ and a PWM modulator. (b) Equivalent block diagram.\label{fig:buck}}
\end{center}
\end{figure*}

Switching converters are integral elements to modern power electronics. Despite their widespread use, they can pose serious challenges to power-supply designers because
almost all of the {\em rules of thumb} governing their design are only applicable to the linearized averaged system even though the system works in switched mode
\cite{erikson}, \cite{lehman}. Switch mode operation is carried out by means of Pulse width Modulation (PWM) action on the switches. In the traditional PWM control, the
duty cycle of the pulse driving signal $\delta(t)$ is varied according to the error $v_e(t)$ between the output voltage and its desired reference. This error is
processed through a compensator to provide the control voltage $v_c(t)$. The simplest analog form of generating a fixed frequency PWM is by comparing the control
voltage with a ramp periodic signal $v_{tri}(t)$ in such a way that the pulse signal goes high/low when the control signal is higher/lower than the triangular signal. A
rising ramp carrier will generate a PWM signal with trailing edge modulation and a falling ramp carrier will generate a PWM signal with leading edge modulation
\cite{lai98}. In the trailing edge modulation strategy, when the control voltage is greater (resp. smaller) than the sawtooth ramp voltage, the switch is ON (resp.
OFF). Whereas in the leading edge modulation strategy, the opposite happens. Figure~\ref{fig:buck} shows a schematic circuit diagram of a buck converter under  voltage
mode control with a compensator $G_c(s)$ and a fixed frequency PWM and its equivalent block diagram.

Probably, the buck converter is the most common voltage regulator in use. Its structure is not complicated (See Fig. \ref{fig:buck}-(b)). The power stage together with
the modulator consists of a pulse generator and a passive low pass filter. For this reason, it can be used to convert an input source voltage $v_g$ into a lower output
voltage $v$. In order to regulate the output voltage, a voltage reference, an error detector, a compensator and a modulator are added to the circuit. These basic
elements form a complete switching buck DC-DC regulator.

In practice, it is desirable that the system operates periodically with a constant switching frequency $f_s=1/T$ equal to that of the external sawtooth ramp  modulating
signal which is the same frequency of the external clock signal. However, under parameter changes, the stability of this operating mode may be lost resulting in
different kinds of instabilities and dynamical behaviors \cite{hamill1}-\cite{bao}. There have been hitherto many research efforts devoted to predict the border of
occurrence of such instabilities. However, in the case of subharmonic oscillations in voltage mode control, most of the reported analysis are based on abstract
mathematical models, which are demonstrated to be very powerful to accurately predict this phenomenon, but not so suited to obtain clear design-oriented criteria in the
parametric design space. In the past studies,  the prediction of subharmonics is mainly based on deriving an accurate discrete time model and then linearizing it in the
vicinity of the operating point. However, the results obtained from such an approach cannot be applied for design purpose. Recently, Filippov's method and the monodromy
matrix were used to predict these instabilities in DC-DC converters and similar results to those obtained from the discrete time approach were derived
\cite{giaouris08}, \cite{gmbpz09}.

\begin{figure*}
\begin{center}
\subfigure[]{\includegraphics[width=6cm]{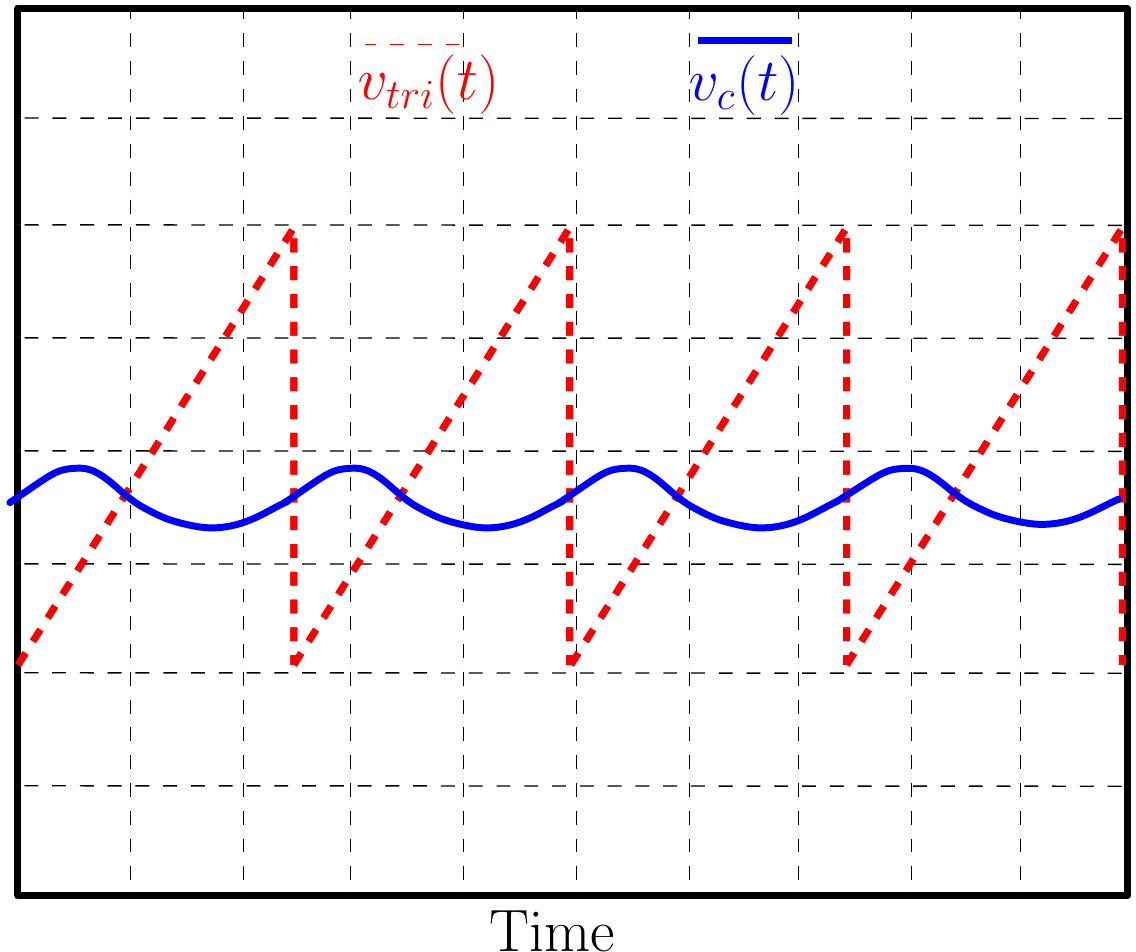}} \subfigure[]{\includegraphics[width=6cm]{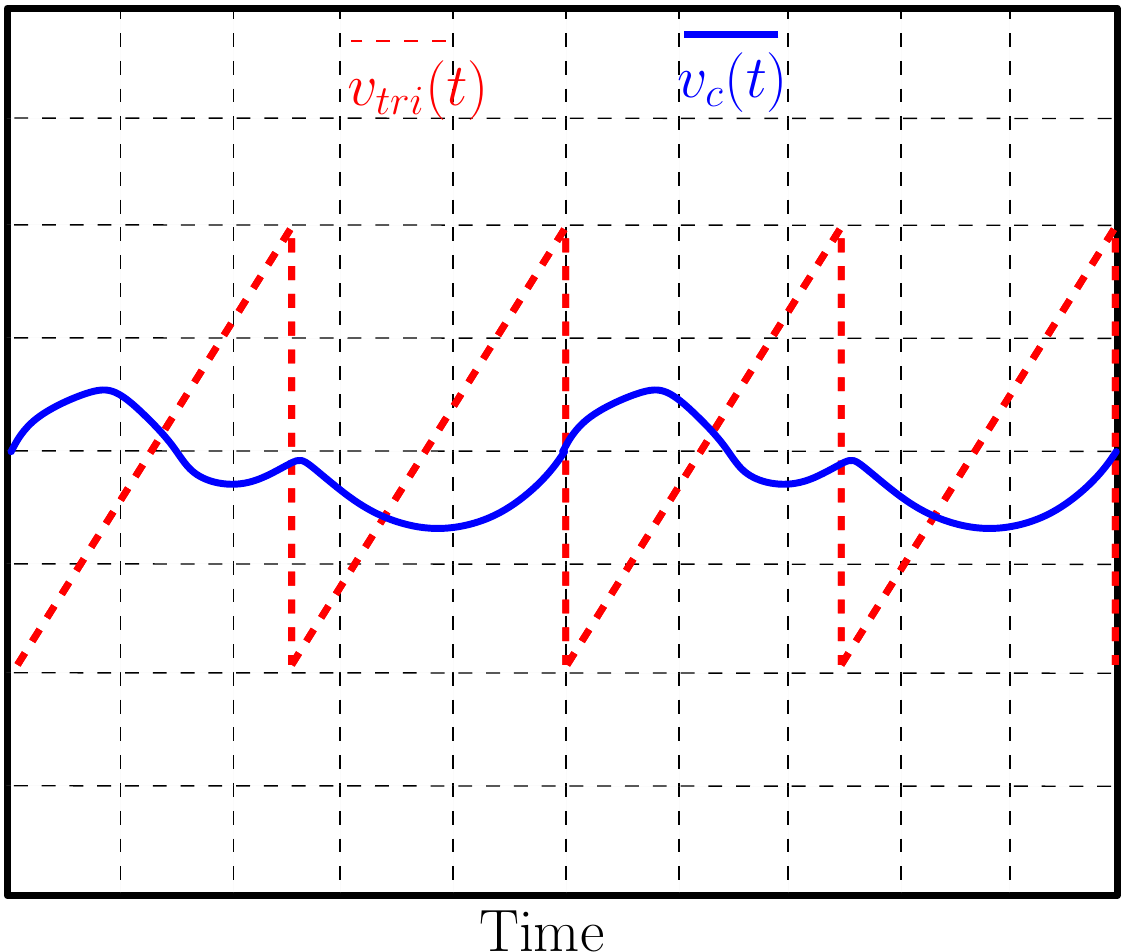}} \subfigure[]{\includegraphics[width=6cm]{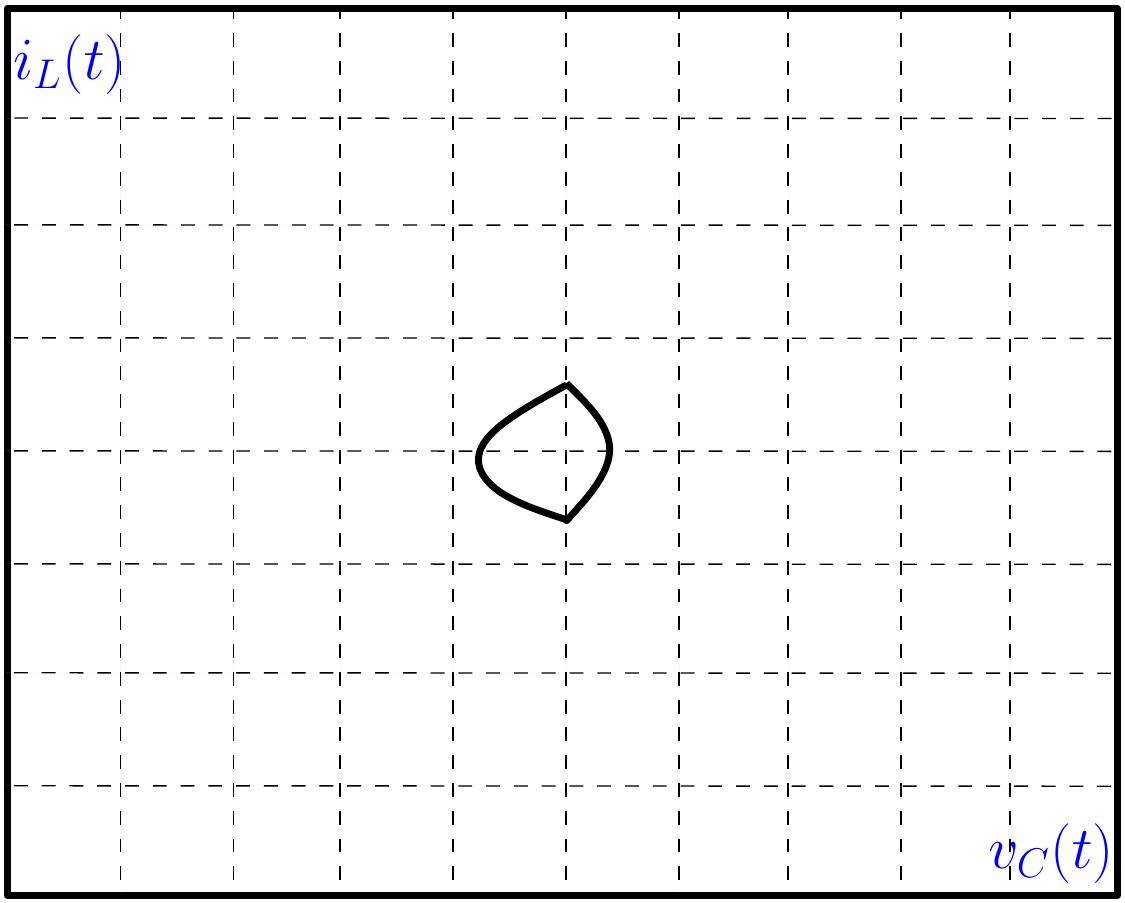}}
\subfigure[]{\includegraphics[width=6cm]{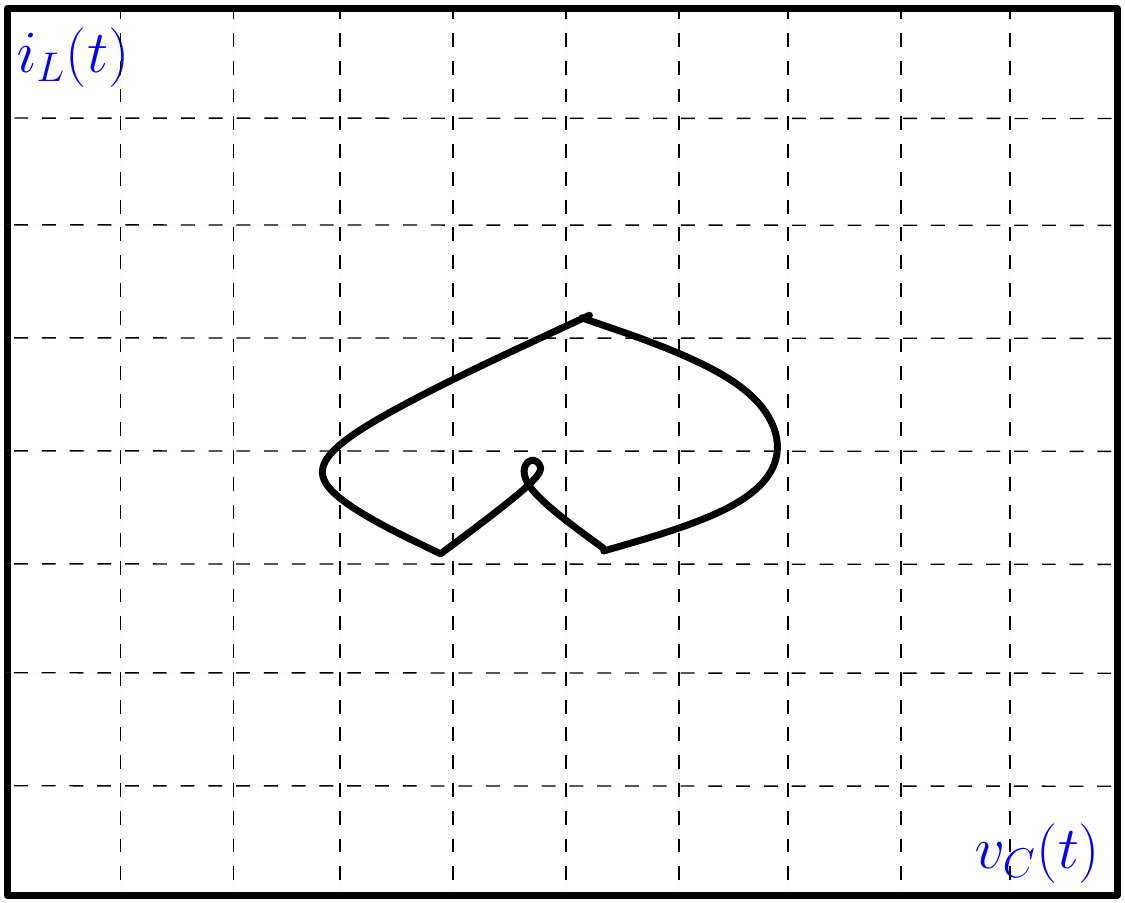}} \subfigure[]{\includegraphics[width=6cm]{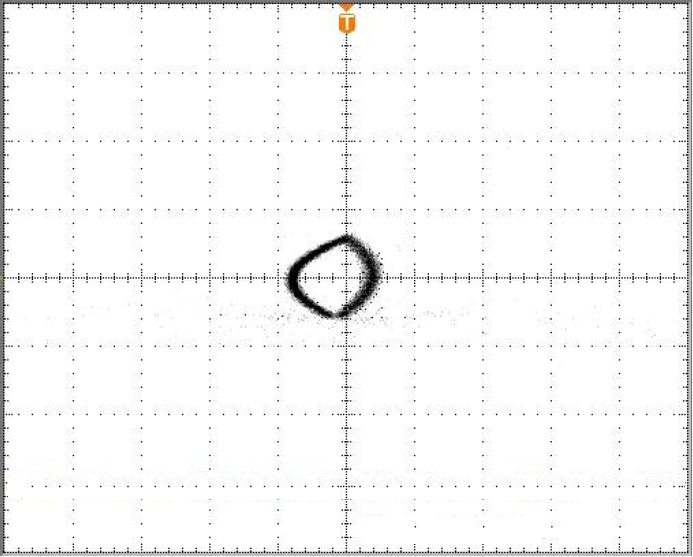}}
\subfigure[]{\includegraphics[width=6cm]{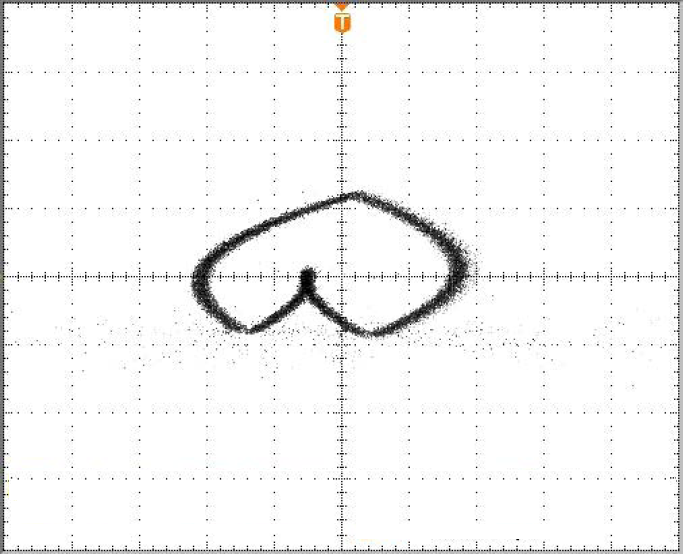}} \caption{Different dynamical behavior of a DC-DC buck converter under voltage mode control. Desired
behavior (left) and subharmonic oscillations (right). (a)-(b) Time domain waveforms of the control voltage $v_c(t)$ and the ramp voltage $v_{tri}(t)$,  (c)-(d)
corresponding state plane trajectories. (e)-(f) corresponding experimental data. \label{fig:dynamics}}
\end{center}
\end{figure*}

While there are powerful design-oriented techniques to characterize low frequency instabilities related to the averaged model \cite{erikson}, there is still a lack of
design oriented tools for predicting fast scale instability in the form of period doubling and subharmonic oscillations inherent to the switched nature of the system.
Note that this phenomenon can be perfectly predicted in the peak and the valley current-mode-controlled system examining mainly the slopes of the inductor current
(\cite{erikson}) or by using modified averaged models taking into account delay terms due to the sample-and-hold effect (\cite{r91}) but not in voltage mode control or
in average current mode control \cite{Tang}, \cite{dixon}. For example, it is known that to guarantee the stability of a peak current-mode-controlled buck converter,
the following inequality, in terms of the duty cycle $D$ and the system parameters,  must hold
\begin{equation}
D-\dfrac{1}{2}<\dfrac{LV_{M}f_s}{v_g}\label{eq:Vmcricmc}
\end{equation}
On the other hand, voltage mode controlled power supply has become very popular especially in low noise next generation communication systems. Such systems are supplied
by means of different Point of Load (PoL) voltage regulators in the form of  single phase voltage-mode-controlled buck converters \cite{panov}-\cite{sun}. It is well
known that these converters, under voltage mode control, are prone to exhibit undesired subharmonic oscillations for certain parameter values \cite{hamill1},
\cite{elaroudi2005}, \cite{diBernardo}. Figure \ref{fig:dynamics} shows typical waveforms and state plane trajectories for the desired periodic behavior and for
subharmonic oscillations in a voltage-mode-controlled buck converter.

An equivalent expression to \eqref{eq:Vmcricmc}, in the case of voltage mode control, will be extremely helpful in designing switching converters free from subhamonic
oscillations. One may, therefore, ask: There exists a similar expression as \eqref{eq:Vmcricmc} for the case of the voltage-mode-controlled system? This paper tries to
develop such an expression for different voltage mode control schemes.

The rest of this paper is organized as follows: a short tutorial on hypergeometric series and their special case, polylogarithm functions, is given in Sections III.
Section IV will revisit the dynamic model of the buck converter in the frequency domain and conditions for periodic behavior and subharmonic oscillations are presented
in this domain by using Fourier series. These series are calculated exactly by using hypergeometric functions and approximately by using polylogarithmic functions. The
approach is then applied, in Sections V, to a voltage-mode-controlled DC-DC buck converter with different controller transfer functions. Some circuit-based simulations
are presented to validate the results obtained from the derived theoretical results. In Section VI, the results are formulated in terms of Figures of Merit, widely used
by the power electronic community, such as the crossover frequency and the phase margin. Finally, in the last section, some concluding remarks of this work are
summarized.

\section{A Short Tutorial on the hypergeometric and polylogarithm functions}
\subsection{Gamma function}
It is somewhat problematic that a large number of definitions have been given for the gamma ($\Gamma$) function \cite{Abramowitz}. Although they describe the same
function, it is not entirely straightforward to prove their equivalence. One of the definitions of the gamma function due to Euler  is \cite{Abramowitz}
\begin{equation}
\Gamma \left( z \right) =\lim_{n\to \infty}\dfrac{n!n^z}{z(z+1)\ldots(z+n)}=\dfrac{1}{z}\prod_{n=1}^\infty\dfrac{1+n}{z+n}
\end{equation}
Using integration by parts, one can show the recursive relation $\Gamma \left( z +1\right) = z\Gamma \left( z \right)$ in such a way that for integer numbers $n$,
$\Gamma \left( n+1\right) =n!$. The Euler gamma function is one of the most important functions in mathematical physics \cite{Abramowitz}. It may be regarded as a
generalization of the factorial which spreads its values over the whole complex plane, except at the negative integers.

\subsection{Psi digamma function}
The psi ($\Psi$) function, also known as the digamma function,  is the function of a complex variable $z$ obtained by differentiating the logarithm of the gamma
function \cite{Abramowitz}, i.e
\begin{equation}
\Psi\left( z \right) = \dfrac{\Gamma'(z)}{\Gamma(z)}
\end{equation}
It is known that $\Psi(1)=-\gamma$, where $\gamma\approx 0.57$ is known as the Euler-Mascheroni constant \cite{Abramowitz}.
\subsection{Pochhammer's Symbol}
The Pochhammer symbol $(z)_n$ is defined as the ratio between $\Gamma(z+n)$ and $\Gamma(z)$ \cite{Abramowitz}, i.e
\begin{equation}
(z)_n = \dfrac{\Gamma(z+n)}{\Gamma(z)}
\end{equation}

\subsection{Generalized hypergeometric functions}
A generalized ($p, q$) order hypergeometric function $\setlength\arraycolsep{1pt} {}_pF_q$ is defined as a power series of the complex number $z$. The expression of
such  function can be written in the form  \cite{Abramowitz}
\begin{eqnarray}
\setlength\arraycolsep{1pt} {}_pF_q\left(\begin{matrix}a_1,& \ldots &a_p\\b_1,& \ldots& b_q\end{matrix};z\right)&=&\sum_{n=0}^{\infty}\dfrac{(a_1)_n(a_2)_n\ldots
(a_p)_n}{(b_1)_n(b_2)_n\ldots (b_q)_n}\dfrac{z^n}{n!}
\label{eq:qFp}
\end{eqnarray}
Many calculus packages such as Mathematica \cite{mathematica}, Maple \cite{maple} and Matlab \cite{matlab} have routines to compute these hypergeometric functions. Yet,
theses functions have interesting properties and in some cases they can be converted to polylogarithms which when evaluated at the unit circle can be expressed by
standard functions \cite{maximon}.

\subsection{Polylogarithm functions}
For some specific values of $a_k$ and $b_k$ in \eqref{eq:qFp}, the hypergeometric functions can be converted to polylogarithm functions \cite{Lewin81}. The
polylogarithm, also known as the Jonqui\'ere's function (See \cite{jonquiere}), is the function defined in the complex plane over the open unit disk by
\begin{equation}
{\rm Li}_n(z)=\sum_{k=1}^{\infty}\dfrac{z^k}{k^n} \label{eq:polylog}
\end{equation}
For $n=1$, one simply obtain ${\rm Li}_1(z)=-\ln(1-z)$. The polylogarithm function is connected to the generalized hypergeometric functions through the relation
\cite{maximon}
\begin{equation*}
{}_{n+1}F_n\left(\begin{matrix}1,a_2& \ldots &a_{n+1}\\a_2+1,a_3+1& \ldots& a_{n+1}+1\end{matrix};z\right)z= {\rm Li}_n(z)
\end{equation*}
More specifically, one has  \cite{maximon}
\begin{eqnarray}
\setlength\arraycolsep{1pt} {}_3F_2\left(\begin{matrix}1,& 1, &1\\2,& 2\end{matrix};z\right)z&=& {\rm Li}_2(z), \quad
\text{and}\label{eq:dilog}\\
\setlength\arraycolsep{1pt} {}_4F_3\left(\begin{matrix}1,& 1, &1, &1\\2,& 2, &2\end{matrix};z\right)z&=& {\rm Li}_3(z) \label{eq:trilog}
\end{eqnarray}
\subsection{Polylogarithm on the unit circle}
For $|z|=1$, $z=e^{j2\pi\Delta}$ ($0<\Delta<1$),  the polylogarithm functions can be expressed as ${\rm Li}_n(e^{j2\pi\Delta})={\rm C}_n(\Delta)+j{\rm S}_n(\Delta)$,
where ${\rm S}_n(\Delta)=\Im[{\rm Li}_n(e^{j\Delta})]$, and
 ${\rm C}_n(\Delta)=\Re[{\rm Li}_n(e^{j\Delta})]$. One has the following first ${\rm C}_n$ and  ${\rm S}_n$
 functions \cite{Abramowitz}
\begin{eqnarray}
{\rm S}_1(\Delta)&=&\pi(\dfrac{1}{2}-\Delta)\label{eq:S1}\\
{\rm C}_2(\Delta)&=&\pi^2(\frac{1}{6}-\Delta+\Delta^2)\label{eq:C2}\\
{\rm S}_3(\Delta)&=&\pi^3\Delta(\frac{1}{3}-\Delta+\frac{2}{3}\Delta^2)\label{eq:S3}
\end{eqnarray}
It will be shown later that \eqref{eq:S1}-\eqref{eq:S3} are the polynomial functions that appear in the expressions establishing the boundary condition for subharmonic
oscillations occurrence in the buck converter. While the first one \eqref{eq:S1} corresponds to the peak and valley current mode control, the second and the third ones
\eqref{eq:C2}-\eqref{eq:S3} appear in both voltage mode control and current mode control with dynamic compensator such as average current mode control \cite{Tang}.

\subsection{The Riemann Zeta function }
It is the function defined by \cite{Abramowitz}
\begin{equation}
\zeta(z)=\sum_{k=1}^{\infty}\dfrac{1}{k^z}\nonumber
\end{equation}
For example, for $z=2$,  $\zeta(2)=\dfrac{\pi^2}{6}$. It can be proved that one has the following values of $\zeta(2k)$ for even integer arguments \cite{Abramowitz}
\begin{eqnarray}
\zeta(4)=\dfrac{\pi^4}{90},\:\: \zeta(6)=\dfrac{\pi^6}{945}\:\: \text{and}\:\:\zeta(2k)=(-1)^{k+1}\dfrac{(2\pi)^{2k}}{2(2k)!}B_{2k}\nonumber
\end{eqnarray}
where $B_k$ are called Bernoulli numbers \cite{sury}. Explicit formulas for the values $\zeta(2k+1)$ for odd integer arguments are not available and currently it seems
that this is a very difficult problem for mathematicians. In fact, it is only proved that $\zeta(3)\approx 1.202$ is an irrational number known as the Ap\'ery's
constant. For engineering use, the fact that $\zeta(k)$ converges quickly to 1 can be used to approximate $\zeta(k)$ by 1 for $k\geq 3$.


\section{Buck converter under voltage mode controller}
The reader may now ask why all previous rather theoretical material is presented? Simply because these functions will appear in the mathematical expressions describing
the periodic behavior and the subharmonic oscillations in the buck converter. The presentation of this theoretical material will allow to the reader a better
understanding of the results presented in this paper.

\subsection{Dynamic model of the power stage circuit}
Consider a buck converter under voltage mode control shown in Fig. \ref{fig:buck}-(a) that can be equivalently represented by the block diagram shown in Fig.
\ref{fig:buck}-(b). For simplicity, let us suppose that $Z_L=Ls+r_\ell$ and $Z_C=r_c+1/Cs$ and that the load is purely resistive ($R$). The approach can be extended to
other more complex models of the loads and reactive components but at the expense of more mathematical involvement. For the considered case, the buck regulator power
stage input-to-output ($\delta$-to-$v$) transfer function $G_{v\delta}(s)$  can be expressed in the following generalized form \cite{erikson}
\begin{equation}
G_{v\delta}(s)=G_{v\delta0}\dfrac{\frac{s}{\omega_{z1}}+1}{\frac{s^2}{\omega_0^2}+\frac{s}{\omega_0Q_0}s+1} \label{eq:buckmodel}
\end{equation}
All the parameters that appear in the expression of the transfer function \eqref{eq:buckmodel} are given in a list of parameters given at the beginning of the this
work. $G_{v\delta0}$ is given by
\begin{equation*}
G_{v\delta0}=\kappa_\ell v_g, \quad\text{where}\quad \kappa_\ell=\dfrac{R}{R+r_\ell}
\end{equation*}
Theoretically, parasitic parameters can be easily ignored. However, the first order parasitic elements such as ESRs of the inductor and the output capacitor are
included in the analysis. The effect of these parasitic elements may be insignificant in some cases but it may be dramatic in others. Engineering judgement can be done
once included to examine their effects.

\subsection{Closed form conditions for periodic behavior and subharmonics}
Let $G_c(s)$ be the compensator transfer function  from the error $v_e(t)$ to the control voltage $v_c(t)$. Let $D$ be the steady state duty cycle and $z=e^{2\pi
j\Delta}$, where $\Delta=D$ for trailing edge modulation and $\Delta=1-D$ for leading edge modulation. Let $V_l$ be the lower value of the triangular ramp modulator
signal $v_{tri}(t)$, $V_M$  its amplitude and  $T$ its period. Let $G(s)=G_{v\delta}(s)G_c(s)$ and $G_k=G(kj\omega_s)$. Let us also define $H_k=G_k/v_g$.

It can be recognized that the voltage at the input of the $Z_\ell Z_cR$ low pass filter is essentially a square wave with amplitude $v_g$. The period of this signal
depends on the dynamics of the closed loop system. Expanding this square wave in a Fourier series and operating on each term $G_k$, equating the resulting expression to
the ramp signal at the switching instants defined by the crossing between $v_c(t)$ and $v_{tri}(t)$, conditions for different periodicity can be obtained. This approach
has been applied in \cite{chb} to obtain the following condition for periodic behavior.
\begin{equation}
v_g(H_0D+2\Re[\sum_{k=1}^{\infty}\dfrac{(1-z^k)H_k}{jk\pi}])-G_{c0}v_{ref}=V_l+\Delta V_M \label{eq:sigm1}
\end{equation}
where $G_{c0}=G_{c}(0)$ and $\Re$ stands for taking the real part. In \cite{chb}, it has been also shown that at  the boundary of subharmonic oscillations, the
following equality holds
\begin{equation}
2v_g\Re[\sum_{k=1}^{\infty}(1-z^k)H_k-H_{k-\frac{1}{2}}]=V_M \label{eq:sigm2}
\end{equation}
The system of equations \eqref{eq:sigm1}-\eqref{eq:sigm2} can be solved for any design parameter to locate the boundary between the desired stable periodic behavior and
subharmonic oscillations. This equation can be solved either graphically by plotting \eqref{eq:sigm1} and \eqref{eq:sigm2} for a sufficiently high number of terms and
looking at the intersection point, by numerical methods or analytically for some limit cases as it will be shown later.

In \cite{chb}, the series in \eqref{eq:sigm1}-\eqref{eq:sigm2} have been approximated by the term that involve the transfer function $G(s)$ with the smallest argument
or solved graphically by truncating the series. In this work, closed form expressions will be given for the series involved in \eqref{eq:sigm1} and \eqref{eq:sigm2}. To
calculate the series, it is required to use the concept of hypergeometric series and polylogarithm functions presented in Section II. It will be shown later that by
considering practical operating conditions, these series can be approximated by standard functions depending on the power stage and controller parameters and more
importantly on the steady state value of the duty cycle $D$.

\subsubsection{Conditions for periodic behavior}
From \eqref{eq:sigm1}, an expression for the input voltage in terms of the circuit parameters and the duty cycle $D$ for the periodic regime is
\begin{equation}
v_{g,1}(D)=\dfrac{G_{c0}v_{ref}+V_l+V_M\Delta}{H_0D+\varepsilon(D)} \label{eq:vg1}
\end{equation}
From \eqref{eq:vg1} and \eqref{eq:sigm1}, $\varepsilon(D)$ is a small quantity depending on $D$ given by
\begin{equation}
\varepsilon(D)=\dfrac{2}{\pi}\Im[\sum_{k=1}^{\infty}\dfrac{(1-z^k)H_k}{k}] \label{eq:epsilon}
\end{equation}
where $\Im$ stands for taking the imaginary part. Without $\varepsilon(D)$, \eqref{eq:vg1} represents the well-known relationship between the input and the output
voltages of the averaged model of a buck converter. Therefore, the $\varepsilon(D)$ term represents a correctional factor that determines the difference between the
averaged model and the reality of the switched system in the case of periodic regime. Note that the same approach is valid for determining other critical parameter
values, such as the feedback gain, ramp amplitude and reference voltage.

\subsubsection{Conditions for subharmonic oscillations}
Similarly, from the equation establishing the subharmonic oscillations boundary \eqref{eq:sigm2}, the input voltage at this boundary is
\begin{equation}
v_{g,2}(D)=\dfrac{V_M}{{\mathcal S}(D)} \label{eq:vg2}
\end{equation}
where from \eqref{eq:sigm2}, ${\mathcal S}(D)$ is given by
\begin{equation}
{\mathcal S}(D)=2\Re[\sum_{k=1}^{\infty}(1-z^k)H_k-H_{k-\frac{1}{2}}] \label{eq:P}
\end{equation}
The function  ${\mathcal S}(D)$ depends on the controller used. In the next sections, closed form expressions will be given for $\varepsilon(D)$ and ${\mathcal S}(D)$
for different  controller transfer functions. These expressions will be derived exactly in terms of hypergeometric functions and approximately by using polylogarithms
and their special expressions on the unit circle using Eqs. \eqref{eq:S1}-\eqref{eq:S3}.

\section{Application to a buck converter under voltage mode control}
\subsection{Buck converter under a simple proportional controller}
It is instructive to start the application of the approach to a buck converter with a simple proportional controller. In this case, the controller transfer function is
simply a constant for all the frequencies ($G_c=k_v$). Both trailing edge modulation and leading edge modulation are studied in a unified approach. The results obtained
for the leading edge modulation case can be applied directly to the trailing edge modulation case by simply changing $D$ by $1-D$.

Let us consider that the output capacitor is ideal. However, the results for this simple case will be valid for a buck converter with a realistic output capacitor and
under a single pole controller with a pole $\omega_{p1}$ located at exactly the zero $\omega_{z1}$ due to the ESR of the output capacitor.

\subsubsection{Condition for periodic oscillations}
By using \eqref{eq:buckmodel} and \eqref{eq:epsilon}, the expression of $\varepsilon(D)$ can be written in the following hypergeometric-function-based form
\begin{eqnarray}
\varepsilon(D)&=&\dfrac{2k_v\kappa_\ell}{\pi}\Im[\sum_{k=1}^{\infty}\dfrac{1-z^k}{k(-\frac{k^2}{\varepsilon_0^2}+j\frac{k}{\varepsilon_0}+1)}]\nonumber\\
&=&\dfrac{k_v\kappa_\ell}{2\pi}\Im[\Psi(1-\varepsilon_d-j\varepsilon_r)+\Psi(1+\varepsilon_d-j\varepsilon_r)\nonumber\\
&-&\dfrac{j}{\sqrt{4Q_0^2-1}}(\Psi(1-\varepsilon_d-j\varepsilon_r)+\Psi(1+\varepsilon_d-j\varepsilon_r))\nonumber\\
&+&\dfrac{2z}{\frac{1}{\varepsilon_0^2}-\frac{j}{\varepsilon_0
Q_0}-1}{}_4F_3\left(\begin{matrix}1,1,1-\varepsilon_d-j\varepsilon_r,1+\varepsilon_d-j\varepsilon_r\\2,2+\varepsilon_d-j\varepsilon_r,2-\varepsilon_d-j\varepsilon_r\end{matrix};
z\right)] \label{eq:epsilonP}
\end{eqnarray}
where $\varepsilon_r=\varepsilon_0/2Q_0$ and $\varepsilon_d=\varepsilon_r\sqrt{4Q_0^2-1}$. For this controller, $G_{c0}=k_v$, $H_0=k_v\kappa_\ell$ and \eqref{eq:vg1}
becomes
\begin{equation}
v_{g,1}(D)=\dfrac{k_vv_{ref}+V_l+V_M\Delta}{k_v\kappa_\ell D+\varepsilon(D)} \label{eq:vg1P}
\end{equation}
Due to the $\omega_{p1}-\omega_{z1}$ pole-zero cancelation, this is the same expression that would be obtained for a single pole controlled buck converter with an ESR
in the output capacitor.

In a practical design, $\omega_0 << \omega_s$ and  $\omega_0Q_0 << \omega_s$ and then, $\varepsilon_0\rightarrow 0$ and $\varepsilon_0Q_0\rightarrow 0$. Then,
\eqref{eq:epsilonP} becomes
\begin{eqnarray}
\lim_{\varepsilon_0\to 0}\lim_{\varepsilon_0Q_0\to
0}\varepsilon(D)=\dfrac{k_v\kappa_\ell\varepsilon_0^2}{\pi}\left(\Im\left[{}_4F_3\left(\begin{matrix}1,1,1,1\\2,2,2\end{matrix}; z\right)z\right]-\zeta(3)\right)
\label{eq:limepsilonP}
\end{eqnarray}
where $\zeta(3)$ is the Ap\'ery's constant (\cite{apery}). After calculating the specific hypergeometric series, \eqref{eq:limepsilonP} gives
\begin{eqnarray}
\lim_{\varepsilon_0\to 0}\lim_{\varepsilon_0Q_0\to 0}\varepsilon(D)&=& \dfrac{k_v\kappa_\ell\varepsilon_0^2}{\pi}\left(\Im\left[{\rm Li}_3(e^{2\pi
j\Delta})\right]-\zeta(3)\right) \label{eq:eD}
\end{eqnarray}
${\rm Li}_3$ is the trilogarithm function defined in general by \eqref{eq:polylog}. The imaginary part of the trilogarithm evaluated at the unit circle is the function
${\rm S}_3$ defined in \eqref{eq:S3} and therefore one has from \eqref{eq:eD}
\begin{eqnarray}
\lim_{\varepsilon_0\to 0}\lim_{\varepsilon_0Q_0\to 0}\varepsilon(D)&=& \dfrac{k_v\kappa_\ell\varepsilon_0^2}{\pi}\left({\rm S}_3(\Delta)-\zeta(3)\right)
\label{eq:vg1P1}
\end{eqnarray}
For design purpose, $\varepsilon(D)$ can even be ignored because the first term in the denominator of \eqref{eq:vg1P} is dominant. In fact it can be proved that
$k_v\kappa_\ell D>>\varepsilon(D)$ for the whole operating  range of the duty cycle $D$ and therefore the input voltage $v_{g,1}(D)$ will be related to the duty cycle
$D$, in the case of a single pole controller canceling the ESR zero $\omega_{z1}$ or in the case of a simple proportional controller, by the following approximated
expression
\begin{equation}
v_{g,1}(D)\approx\dfrac{k_vv_{ref}+V_l+V_M\Delta}{k_v\kappa_\ell D} \label{eq:vg11}
\end{equation}
Note that this is the same expression that one would obtain if a simple averaged model is used. For design-oriented analysis \eqref{eq:vg11} is accurate  enough while
an approximated expression for $\varepsilon(D)$ is still available in terms of standard functions. Indeed, if ${\rm S}_3$ is substituted by its expression in
\eqref{eq:S3} and by using \eqref{eq:vg1P}, another approximated closed form expression for $v_{g,1}(D)$ is
\begin{equation}
v_{g,1}(D)\approx \dfrac{3(k_vv_{ref}+V_l+V_M\Delta)}{3k_v\kappa_\ell D+k_v\kappa_\ell\varepsilon_0^2\pi^3\Delta(1-3\Delta+2\Delta^2)} \label{eq:vg1Psimple}
\end{equation}
Note that both \eqref{eq:vg11} and \eqref{eq:vg1Psimple} are approximated expressions while the exact equation is \eqref{eq:vg1P}.
\subsubsection{Condition for subharmonic oscillations}
Following a similar procedure to that of the previous section, an expression for ${\mathcal S}(D)$ in \eqref{eq:P} can be obtained. For the case of single pole
controller canceling the zero $\omega_{z1}$ due to the ESR of the output capacitor or in the case of a simple proportional controller and an ideal output capacitor,
from \eqref{eq:buckmodel} and \eqref{eq:vg2}$, {\mathcal S}(D)$ is given by the following hypergeometric-function-based expression
\begin{eqnarray}
{\mathcal S}_{\rm
p}(D)&=&2k_v\kappa_\ell\Re[\dfrac{1}{\sqrt{4Q_0^2-1}}(\varepsilon_0Q_0(\Psi(1-\varepsilon_d-j\varepsilon_r)-\Psi(\frac{1}{2}+\varepsilon_d+j\varepsilon_r))\nonumber\\
&-&\Psi(1+\varepsilon_d-j\varepsilon_r)+\Psi(\frac{1}{2}+\varepsilon_d-j\varepsilon_r))\nonumber\\
&-&\dfrac{z}{\frac{1}{\varepsilon_0^2}-\frac{j}{\varepsilon_0 Q_0}-1
}{}_3F_2\left(\begin{matrix}1,1+\varepsilon_d-j\varepsilon_r,1-\varepsilon_d-j\varepsilon_r\\2-\varepsilon_d-j\varepsilon_r,2+\varepsilon_d-j\varepsilon_r\end{matrix};
z\right)] \label{eq:Pp}
\end{eqnarray}
Under the practical conditions, $\varepsilon_0\rightarrow 0$ and $\varepsilon_0Q_0\rightarrow 0$, \eqref{eq:Pp} becomes
\begin{eqnarray}
\lim_{\varepsilon_0\to 0}\lim_{\varepsilon_0Q_0\to 0}{\mathcal S}_{\rm
p}(D)&=&k_v\varepsilon_0^2\left(2\zeta(2)+\Re\left[{}_3F_2\left(\begin{matrix}1,1,1\\2,2\end{matrix}; z\right)z\right]\right) \label{eq:3F2}
\end{eqnarray}
By calculating the special hypergeometric series that appears in \eqref{eq:3F2}, ${\mathcal S}_{\rm p}(D)$ can be written in the following dilogarithm-based expression
\begin{eqnarray}
\lim_{\varepsilon_0\to 0}\lim_{\varepsilon_0Q_0\to 0}{\mathcal S}_{\rm p}(D)&=& k_v\kappa_\ell\varepsilon_0^2(2\zeta(2)+\Re[{\rm Li}_2(e^{2\pi j \Delta})])
\label{eq:vg2dilog}
\end{eqnarray}
Using \eqref{eq:vg2} and \eqref{eq:vg2dilog}, the critical input voltage $v_{g,2}(D)$ giving the boundary of subharmonic oscillation in terms of the duty cycle $D$ is
obtained
\begin{equation}
v_{g,2}(D) \approx\dfrac{V_M}{k_v\kappa_\ell\varepsilon_0^2(2\zeta(2)+\Re[{\rm Li}_2(e^{2\pi j \Delta})])}
\end{equation}
Taking the real part of the dilogarithm function using \eqref{eq:C2} and substituting $\Delta$, the following expression is obtained for both the leading edge
modulation and the trailing edge modulation modulation strategies
\begin{eqnarray}
v_{g,2}(D) &\approx& \dfrac{V_M}{k_v\kappa_\ell\varepsilon_0^2}\dfrac{1}{\pi^2(1-2D(1-D))} \label{eq:vg2DbarD}
\end{eqnarray}
which is the same expression obtained in \cite{erla10} for the case of ideal components ($r_c=0$ and $r_\ell =0$) by using a very different approach. Note that
$\kappa_\ell$ and $\varepsilon_0$ in \eqref{eq:vg2DbarD} depends on the parasitic parameters.  The critical value of the input voltage depends on both power stage and
control system parameters and most importantly on the duty cycle $D$.

%

From \eqref{eq:vg2DbarD}, subharmonic oscillations can be avoided in the system if the following inequality holds
\begin{eqnarray}
\underbrace{\dfrac{1}{2}-D+D^2}_{f(D)}<\underbrace{\dfrac{V_M\omega_s^2}{2v_gk_v\kappa_\ell\omega_0^2\pi^2}}_{g(p)} \label{eq:polysub}
\end{eqnarray}
where $f$ is a second degree polynomial function of the duty cycle $D$ and $g$ is a function of the vector $p$ of the system  parameters. Eq.~\eqref{eq:polysub} can be
considered as an extended expression to \eqref{eq:Vmcricmc} for the voltage-mode-controlled buck converter. It can be observed that \eqref{eq:polysub} is invariant
under the change $D\rightarrow 1-D$ which implies that the same expression is valid  for both leading edge modulation and trailing edge modulation strategies. To
validate the previous theoretical predictions, let us consider the following example.

\begin{example}
Consider the well known and widely studied example of  buck converter considered first in \cite{hamill1} and later by other researchers in \cite{TSE03},
\cite{diBernardo}, \cite{fossas}, \cite{giaouris08}. The same set of parameter values will be considered so that the readers can make the comparison easily. Namely,
$L=20$~mH, $C=47$~$\mu$F, $V_l=3.8$~V, $V_M=4.4$~V, $T=400$~ $\mu$s, $v_{ref}=11.3$~V and $k_v=8.4$. Furthermore, numerical simulations are not repeated here to save
space. Interested readers can see \cite{giaouris08} and \cite{gmbpz09} for both numerical simulations and experimental measurements.
\begin{figure*}
\begin{center}
\subfigure[$v_{g,1}(D)$ and $v_{g,2}(D)$ in dB]{\includegraphics[width=8cm]{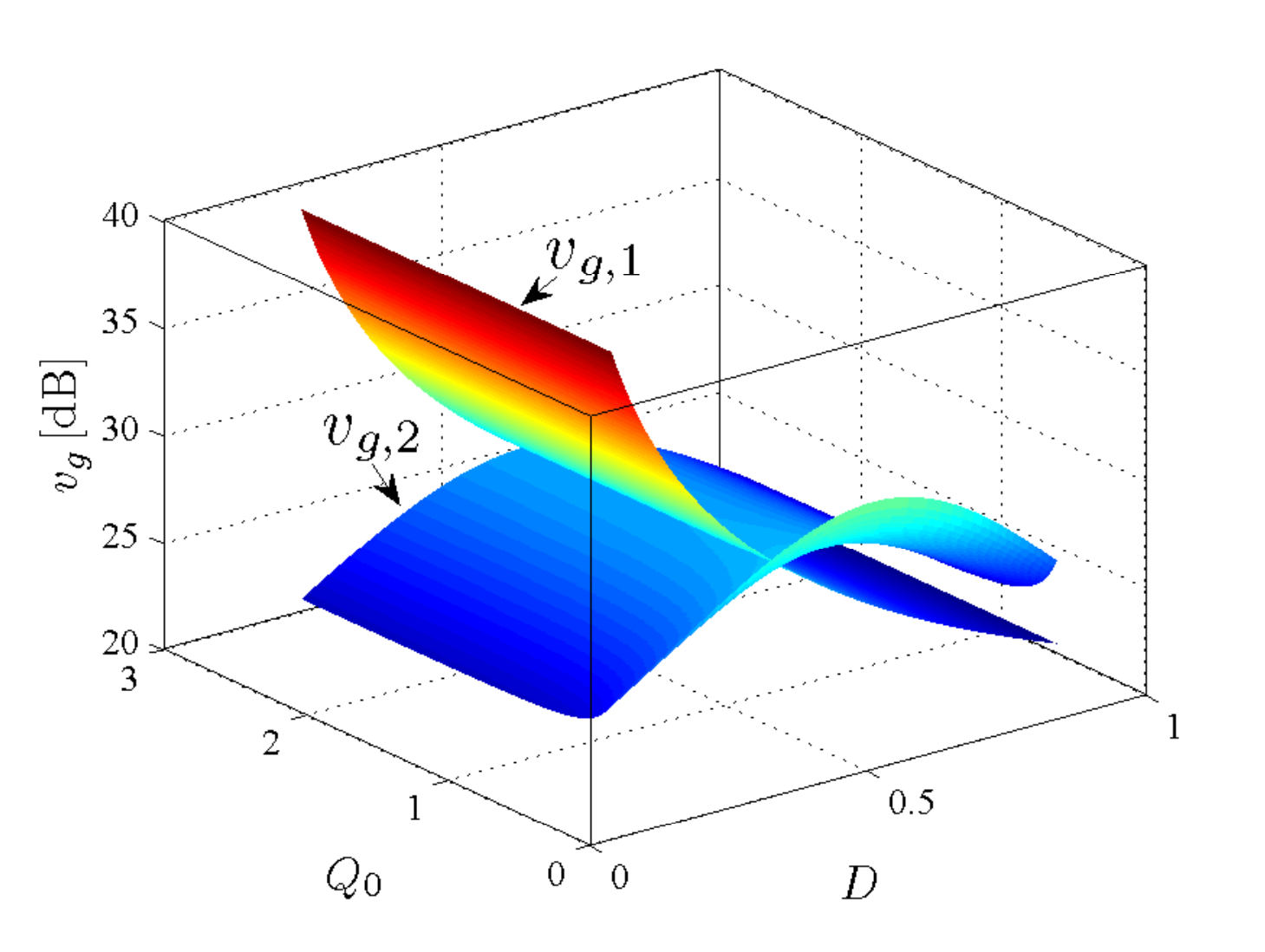}} \subfigure[Exact and approximated
$v_{g,2}(D)$]{\includegraphics[width=8cm]{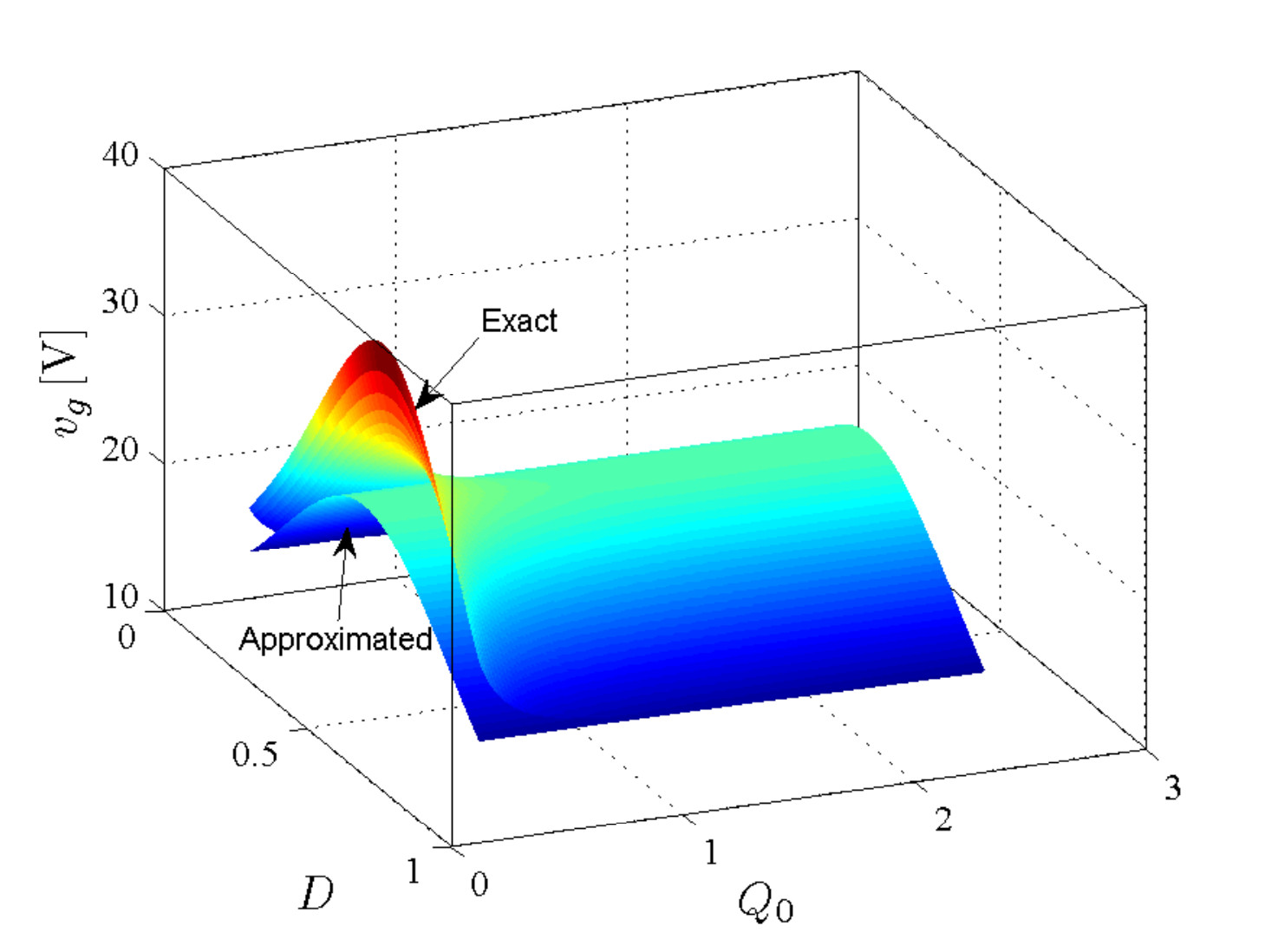}} \caption{(a) Mesh of $v_{g,1}(D)$ and $v_{g,2}(D)$ in terms of the duty cycle $D$ and the quality factor $Q_0$
where $v_{g,1}(D)$ and $v_{g,2}(D)$ are represented in dB.  (b) Mesh of $v_{g,2}(D)$ from \eqref{eq:vg2} and \eqref{eq:vg2DbarD} showing that for small $Q_0$,
\eqref{eq:vg2DbarD} is not enough accurate. \label{fig:vg1vg2}}
\end{center}
\end{figure*}

Figure \ref{fig:vg1vg2}-(a) shows a mesh plot of $v_{g,2}(D)$ from \eqref{eq:vg2}  and $v_{g,1}(D)$ from \eqref{eq:vg1} in terms of the duty cycle $D$ and the quality
factor $Q_0$ or equivalently the load resistance $R$. The intersection of the two surfaces is the curve of the critical input voltage at the boundary of subharmonic
oscillations. In Fig. \ref{fig:vg1vg2}-(b), the exact mesh plot of $v_{g,2}(D)$ from \eqref{eq:vg2} is shown  together with the approximated plot from
\eqref{eq:vg2DbarD}. From this figure, it can be observed that for small $Q_0$, a discrepancy exists between the exact and the approximated plots. This discrepancy
becomes significant for a power quality $Q_0$ approaching $1/2$. For $Q_0<1/2$, \eqref{eq:vg2DbarD} will give inaccurate results. In fact, the main drawback of the
simplified expression based on \eqref{eq:vg2DbarD} is that it practically does not depend on the load resistance $R$ which is an important design
parameter\footnote{Note that the unique dependence of the simplified expression of $v_{g,2}(D)$ on $R$ is through the parameter $\kappa_\ell\approx 1$.}.

\begin{figure*}
\begin{center}
\subfigure[$R=5\:\:\Omega\Rightarrow Q_0\approx 0.24$]{\includegraphics[width=8cm]{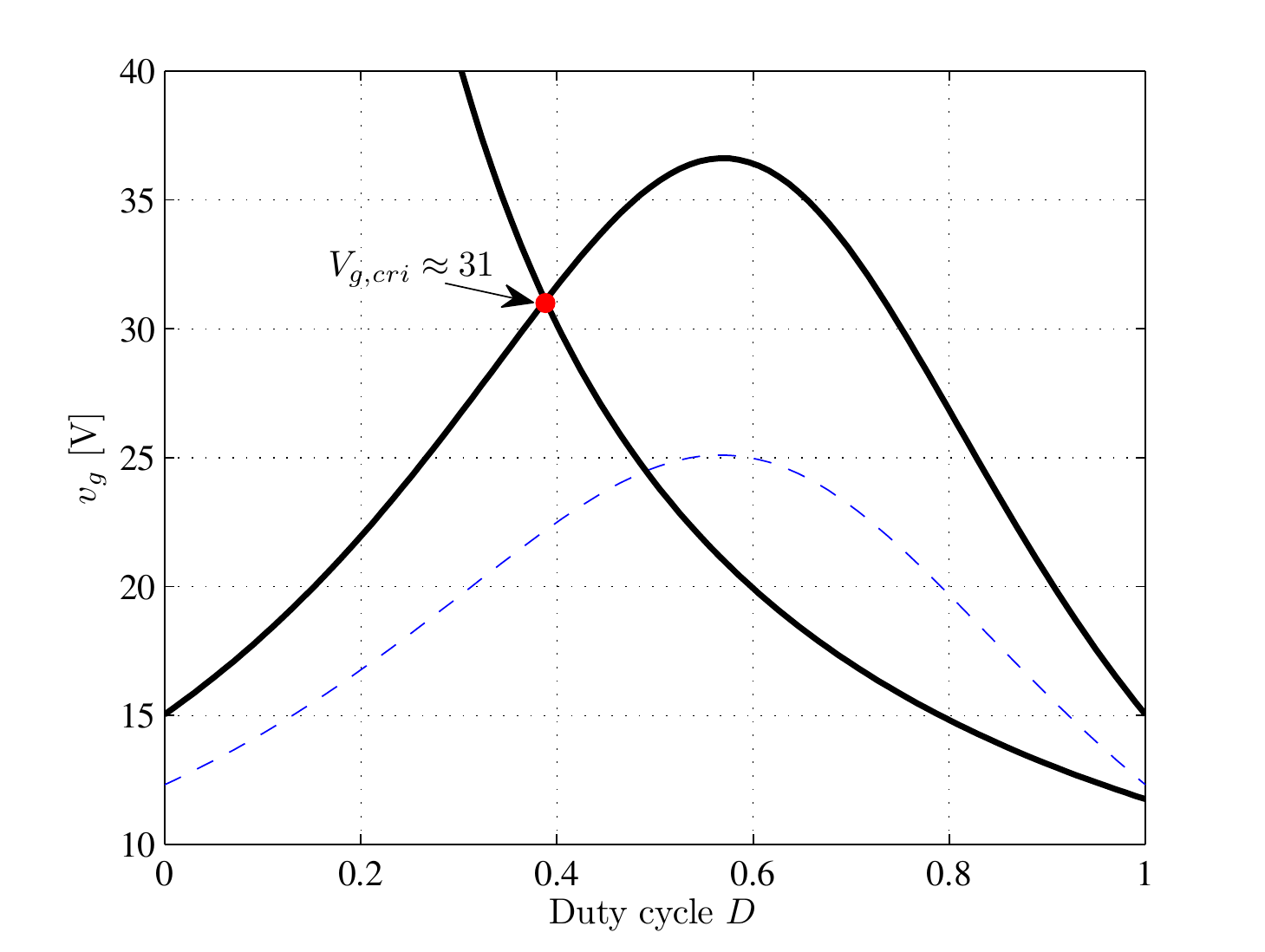}} \subfigure[$R=12\:\:\Omega\Rightarrow Q_0\approx
0.58$]{\includegraphics[width=8cm]{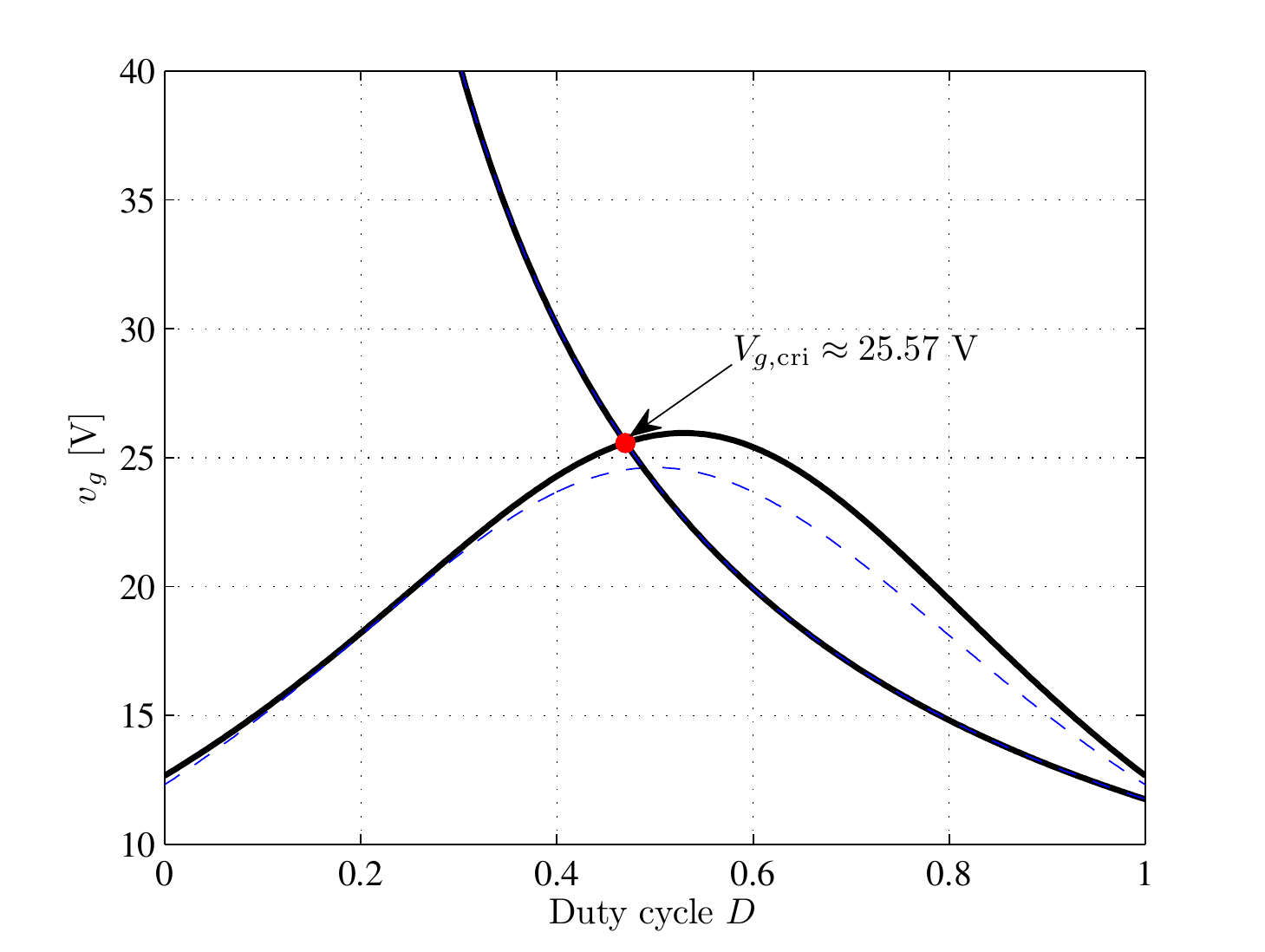}} \subfigure[$R=22\:\:\Omega\Rightarrow Q_0\approx 1$]{\includegraphics[width=8cm]{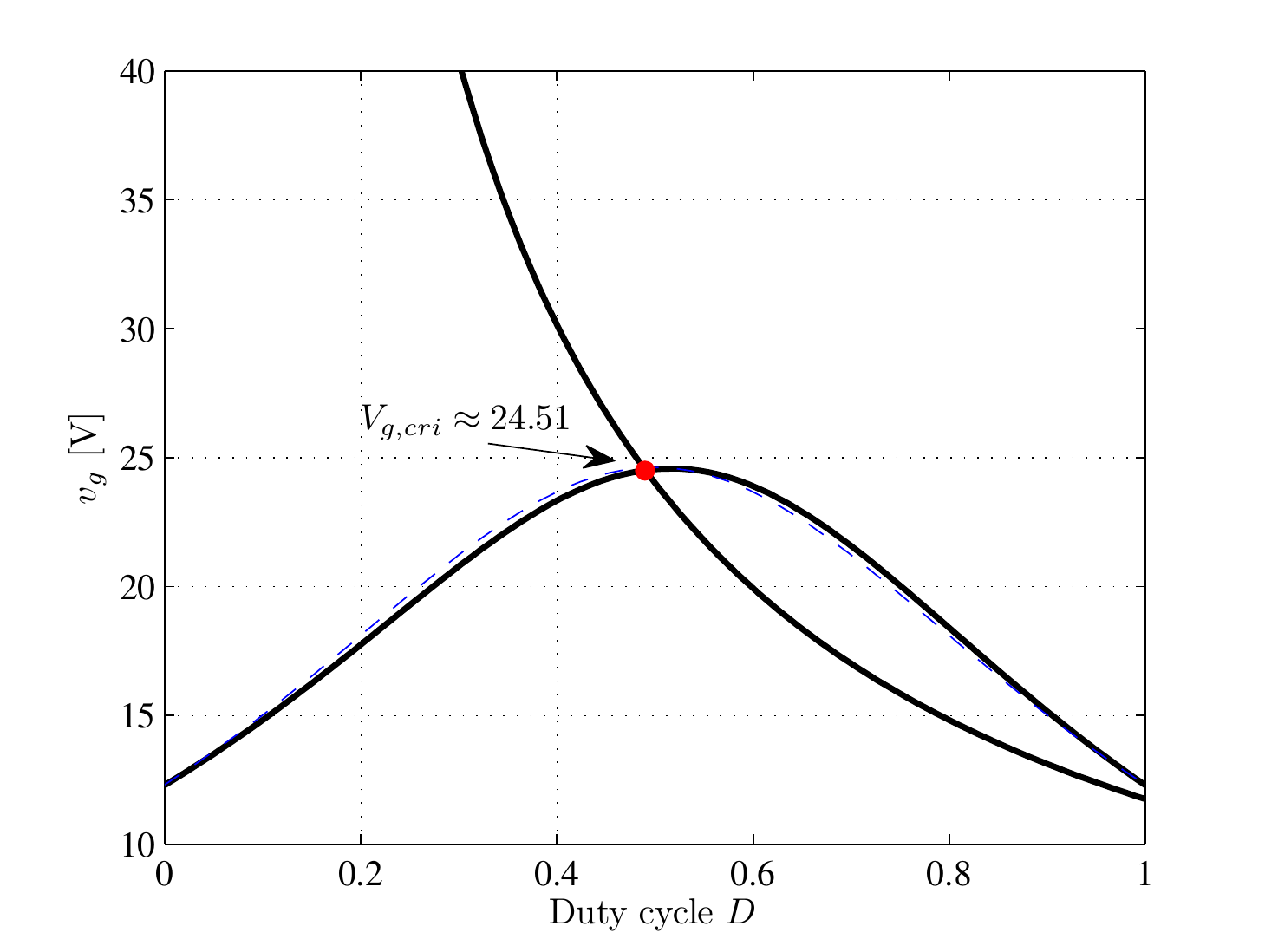}}
\subfigure[$R=50\:\:\Omega\Rightarrow Q_0\approx 2.42$]{\includegraphics[width=8cm]{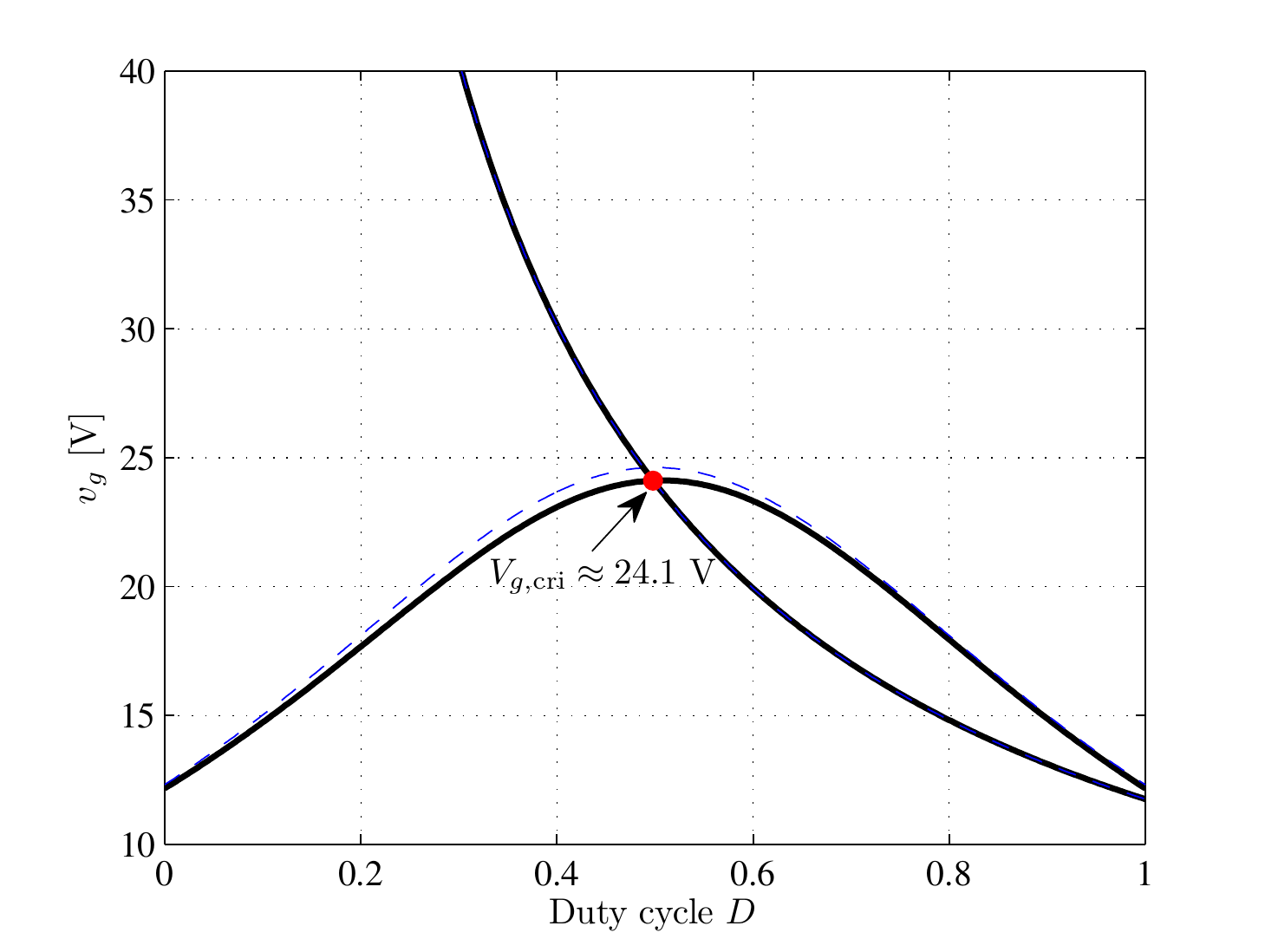}} \caption{Stability boundaries of the buck converter under a simple
proportional control in terms of the duty cycle for different values quality factor $Q_0$ or equivalently the load resistance $R$. Solid: exact curves. Dashed:
approximated curves. \label{fig:vg1vg22D}}
\end{center}
\end{figure*}

Figure \ref{fig:vg1vg22D} shows the exact plots of $v_{g,2}(D)$ and $v_{g,1}(D)$ from \eqref{eq:vg1} and \eqref{eq:vg2} and the approximated plots from
\eqref{eq:vg1Psimple} and \eqref{eq:vg2DbarD} for different values of the load resistance $R$. For all the considered values of this parameter, the curves of
$v_{g1}(D)$ using \eqref{eq:vg1} and \eqref{eq:vg1Psimple} are practically coincident. However, concerning $v_{g,2}(D)$, it can be observed that there is a discrepancy
between the exact curve and the approximated one for low values of the load resistance.

The critical value of the input voltage $V_{g,{\rm cri}}$ is the intersection of the curves $v_{g,1}(D)$ from \eqref{eq:vg1} and $v_{g,2}(D)$ from \eqref{eq:vg2}. It
can be determined by simply subtracting $v_{g,1}(D)$ from $v_{g,2}(D)$, equating the resulting expression to zero and solving for $D$ and then substituting in
$v_{g,1}(D)$ or in $v_{g,2}(D)$. For instance, for $R=22$ (Fig. \ref{fig:vg1vg22D}-(c)), the critical value of the input voltage $V_{g,{\rm cri}}\approx 24.51$ V (for
$D\approx 0.47$) which is in perfect agreement with \cite{diBernardo}, \cite{giaouris08}, \cite{gmbpz09}. For this value of the load resistance, $Q_0\approx 1$ and the
critical value can be also obtained accurately from the approximated expression.

As $Q_0$ decreases, the value of intersection point $V_{g,{\rm cri}}$ increases as predicted in \cite{giaouris08} by using a different approach based on Fillipov method
and the monodromy matrix. For example, for $R=$ 5 $\Omega$, the critical value of the input voltage is approximately 31 V which is in perfect agreement with
\cite{giaouris08} (See Fig. 7 in \cite{giaouris08}). From the approximated expression of $v_{g,2}(D)$ one still obtains $V_{g,{\rm cri}}\approx 24.51$ V. This
discrepancy is due to the low value of $Q_0$.

Note that the symmetry with respect to $D=1/2$ is lost for low values of the quality factor $Q_0$ which does not appear in \eqref{eq:vg2DbarD}. Also, the maximum value
of $v_{g,2}(D)$ moves to the right side of $D=1/2$, when $Q_0$ decreases in the case of leading edge modulation while it moves to the left side in the case of trailing
edge modulation.
\end{example}

\subsection{Buck converter under a PI controller}
A simple dynamic compensator that remove the steady state error from the system under a proportional control, can be obtained by adding a pole at the origin. Also a
zero $\omega_{z2}$ is added to improve the phase margin. The resulting controller is a Proportional-Integral  (PI) compensator and generally has the following transfer
function
\begin{equation}
G_c(s)=k_v\dfrac{s+\omega_{z2}}{s}
\end{equation}
The added zero $\omega_{z2}$ is selected to be at $\alpha\omega_0$ ($0.5<\alpha<1.6$) \cite{Hegarty}, \cite{Day}. Note that the total loop transfer function will be the
same one corresponding to a Type II controller with an aditional pole $\omega_{p1}$ canceling the zero $\omega_{z1}$ due to the ESR of the output capacitor \cite{Day}.
The exact expression of $v_{g,1}(D)$ can also be expressed by \eqref{eq:vg1}, where $\varepsilon(D)$ is also given by a generalized hypergeometric function and can be
approximated by polylogarithms. However, as $G_{c0}$ and $H_0$ are infinite for this type of controllers due to the pole at the origin, from \eqref{eq:vg1},
$v_{g,1}(D)$ can be approximated by $v_{ref}/D$ which is a very simple and well known expression for $v_{g,1}(D)$ relating the input and the output voltages in a buck
converter that can also be obtained from a conventional averaged model \cite{erikson}. Following a similar procedure as for the case of a simple proportional
controller, the exact expression of $v_{g,2}(D)$ in this case is also given by a hypergeometric function. The same expression of \eqref{eq:vg2} is valid in this case
with ${\mathcal S}_{\rm pi}$ depending on a new parameter $\varepsilon_{\alpha}=\alpha\varepsilon_0$ and given by the following expression
\begin{eqnarray}
{\mathcal S}_{\rm pi}(D)&=&\dfrac{k_v\kappa_\ell}{2}\Re[\dfrac{1}{\sqrt{4Q_0^2-1}}(\Psi(1-\varepsilon_d-j\varepsilon_r)\varepsilon_\alpha-2\varepsilon_0Q_0\Psi(1-\varepsilon_d-j\varepsilon_r)\nonumber\\
&+&\Psi(1+\varepsilon_d-j\varepsilon_r)\varepsilon_\alpha-2Q_0\varepsilon_0\Psi(1+\varepsilon_d-j\varepsilon_r)\nonumber\\
&+&\Psi(\frac{1}{2}-\varepsilon_d-j\varepsilon_r)\varepsilon_\alpha-2\varepsilon_0Q_0\Psi(\frac{1}{2}-\varepsilon_d-j\varepsilon_r)\nonumber\\
&-&\Psi(\frac{1}{2}+\varepsilon_d-j\varepsilon_r)\varepsilon_\alpha+2\varepsilon_0Q_0\Psi(\frac{1}{2}+\varepsilon_d-j\varepsilon_r))\nonumber\\
&+&j\Psi(\frac{1}{2}+\varepsilon_d-j\varepsilon_r)\varepsilon_\alpha+j\Psi(\frac{1}{2}-\varepsilon_d-j\varepsilon_r)\varepsilon_\alpha\nonumber\\
&+&j\Psi(1+\varepsilon_d-j\varepsilon_r)\alpha\nonumber-j\Psi(1-\varepsilon_d-j\varepsilon_r)\alpha\nonumber\\
&+&\dfrac{2jz(j+\varepsilon_\alpha)}{1-\frac{1}{\varepsilon_0^2}+j\frac{1}{\varepsilon_0Q_0}} {}_5F_4\left(\begin{matrix}
1, 1, 2-j\varepsilon_\alpha, 1-\varepsilon_d-j\varepsilon_r, 1+\varepsilon_d-j\varepsilon_r\\
2, 1-j\varepsilon_\alpha, 2-\varepsilon_d-j\varepsilon_r, 2+\varepsilon_d-j\varepsilon_r\end{matrix}; z\right)] \label{eq:Ppi}
\end{eqnarray}
Note that when $\alpha\rightarrow 0$, $\varepsilon_\alpha\rightarrow 0$, and ${\mathcal S}_{\rm pi}(D)\rightarrow {\mathcal S}_{\rm p}(D)$. To obtain a more simple
expression for $v_{g,2}(D)$ in the case of a type II and PI controllers, let us consider $\varepsilon_0 \rightarrow 0$ and $\varepsilon_0Q_0 \rightarrow 0$, which is
the case in almost all practical designs. Therefore, in terms of polylogarithms, \eqref{eq:Ppi} becomes  as follows
\begin{eqnarray}
\lim_{\varepsilon_0\to 0}\lim_{\varepsilon_0Q_0\to 0}{\mathcal S}_{\rm pi}(D)&=&k_v\kappa_\ell\varepsilon_0^2(\Re[{\rm Li}_2(e^{2\pi
j\Delta})]+\zeta(2))-\alpha\varepsilon_0^3\Im[{\rm Li}_3(e^{2\pi j\Delta})]
\end{eqnarray}
It can be noted that, as $\omega_s>>\omega_0$ ($\varepsilon_0\rightarrow 0$), the contribution due to the trilogarithm term is actually negligible. The real part of the
dilogarithm and the imaginary part of the trilogarithm have been already obtained before. Therefore, for the case of Type II and the PI controllers with a trailing edge
modulation ($\Delta=D$), the approximated critical value of the feedback gain has the following expression
\begin{eqnarray}
k_{v}(D) &\approx& \dfrac{3V_M}{\kappa_\ell v_g\varepsilon_0^2}\dfrac{1}{3\pi^2(1-2D(1-D))+\alpha\varepsilon_0\pi^3D(1-3D+2D^2)} \label{eq:vg2DbarDpi}
\end{eqnarray}
Similar analysis can be applied to obtain the expression corresponding to a leading edge modulation strategy. Note that the symmetry with respect to $D=1/2$ is lost
although this cannot be appreciated since $\varepsilon_0$ is very small in practice. Equation \eqref{eq:vg2DbarDpi} can be considered as an extension of
\eqref{eq:vg2DbarD} for the case of the Type II and the PI controllers. The next example validates the results for a buck converter  for these two different
controllers.

\begin{example}
Consider a buck converter with voltage mode control and trailing edge modulation studied in \cite{erla10} whose value of parameters are oriented to miniaturization. The
considered fixed parameters in this example are: $R=1$~$\Omega$, $L = 30$~nH, $C=50$~nF, $f_s=50$~MHz, $V_l=$ 0, $V_M=$~1 V, $v_{g}=3$ V and $\alpha=1/2$. Two cases
will be studied. In the first one, ideal output capacitor and inductor are considered and the system will be under a PI control. In the second case, parasitic elements
will be included. While for the ESR $r_\ell$ of the inductor, only static performances are changed, in the case of an ESR $r_c$ of the output capacitor, a zero
$\omega_{z1}$ is introduced in the dynamic model of the system. This zero is canceled with an additional $\omega_{p1}$ of the type II controller. Note that although the
dynamic effect of the ESR output capacitor zero is canceled, its static effect still appears in the expression of $v_{g,2}(D)$ since $\varepsilon_0$ depends on this
parameter.

\begin{figure*}[h!]
\begin{center}
\subfigure[$r_c~=~0$, $r_\ell=0$, $\alpha=1/2$]{\includegraphics[width=8cm]{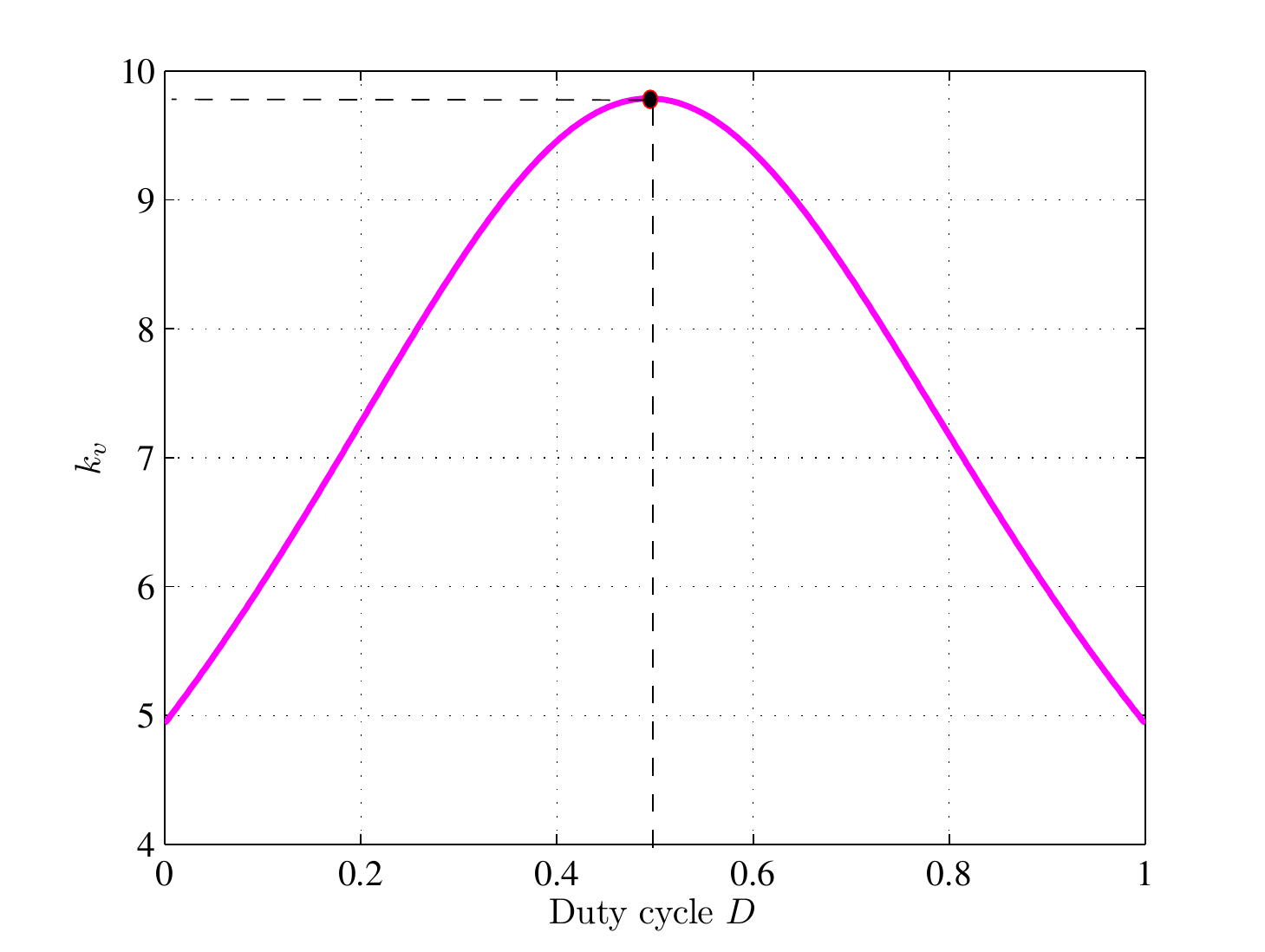}} \subfigure[$r_c~=~0.05\:\: \Omega$,  $r_\ell~=~0.1$ $\Omega$,
$\alpha=1/2$]{\includegraphics[width=8cm]{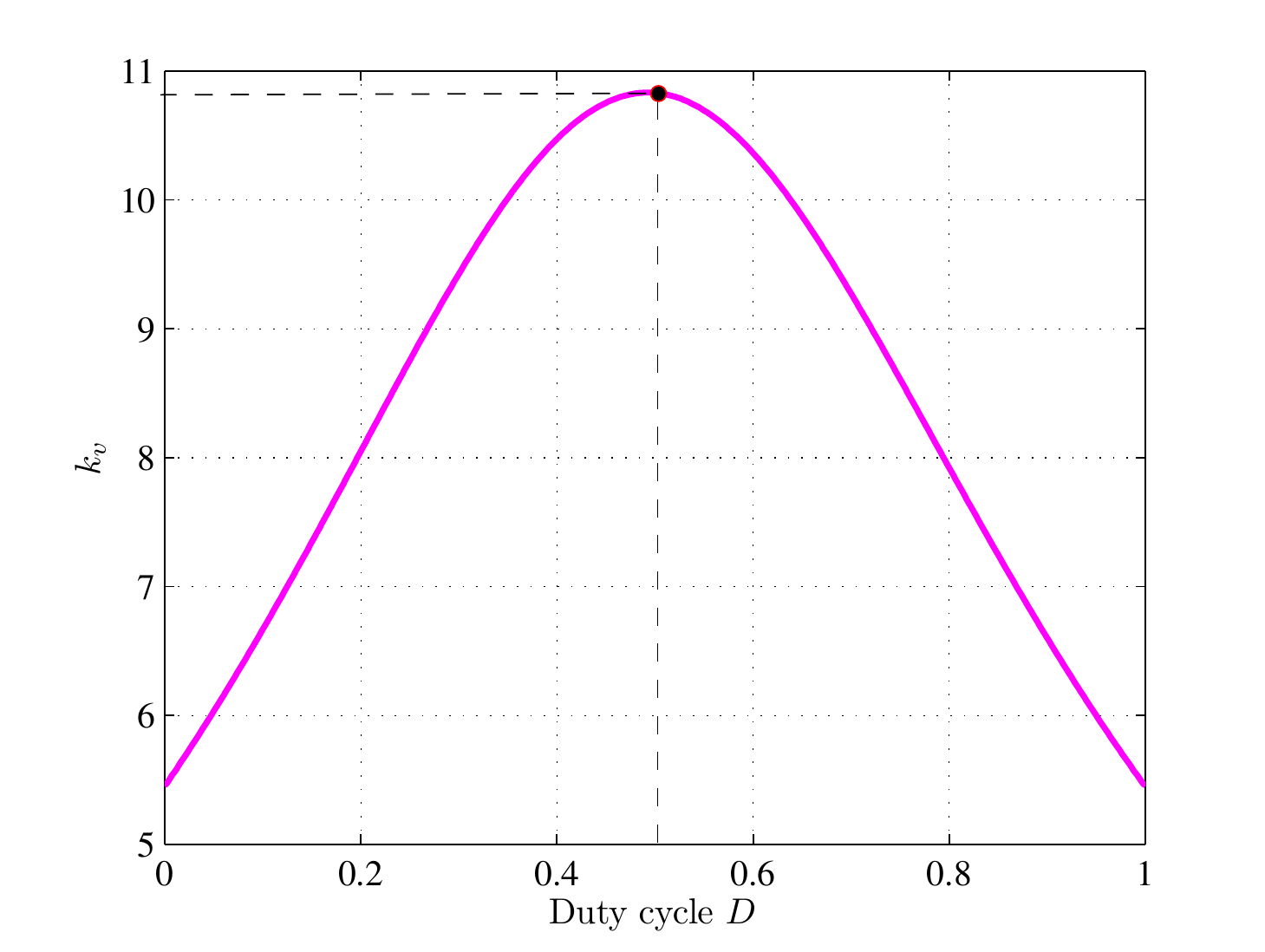}} \caption{Stability curves of the buck converter of Example 2. (a) $k_v$ in terms of the duty cycle $D$ for an
ideal converter $k_{v, {\rm cri}}\approx 9.7$. (b) $k_v$ in terms of the duty cycle $D$ taking into account parasitic elements. $k_{v, {\rm cri}}\approx 10.7$.
\label{fig:vg1vg2PI}}
\end{center}
\end{figure*}

\subsubsection{Ideal components}
The system in this case is under a PI controller. Furthermore, in order to do not induce slow scale instability, let us select $\alpha$ such that the zero
$\omega_{z2}=\omega_0/2$ ($\alpha=1/2$) of the PI controller be below $\omega_0Q_0$ \cite{erla10}. Therefore, it will be below $\omega_s$ because $\omega_s>\omega_0Q_0$
in a practical design. Figure \ref{fig:vg1vg2PI}-a shows the stability curve \eqref{eq:vg2DbarDpi} in terms of the proportional gain $k_v$ and the duty cycle $D$. From
\eqref{eq:vg2DbarDpi}, the critical value of the feedback gain for the previous set of parameter values with $v_{ref}=~1.5$~V ($D=1/2$), is $k_{v, {\rm cri}}\approx
9.7$.

\begin{figure*}[h!]
\begin{center}
\subfigure[$r_c~=~0$, $r_\ell=0$, $\alpha=1/2$, $k_v=9.5$]{\includegraphics[width=8cm]{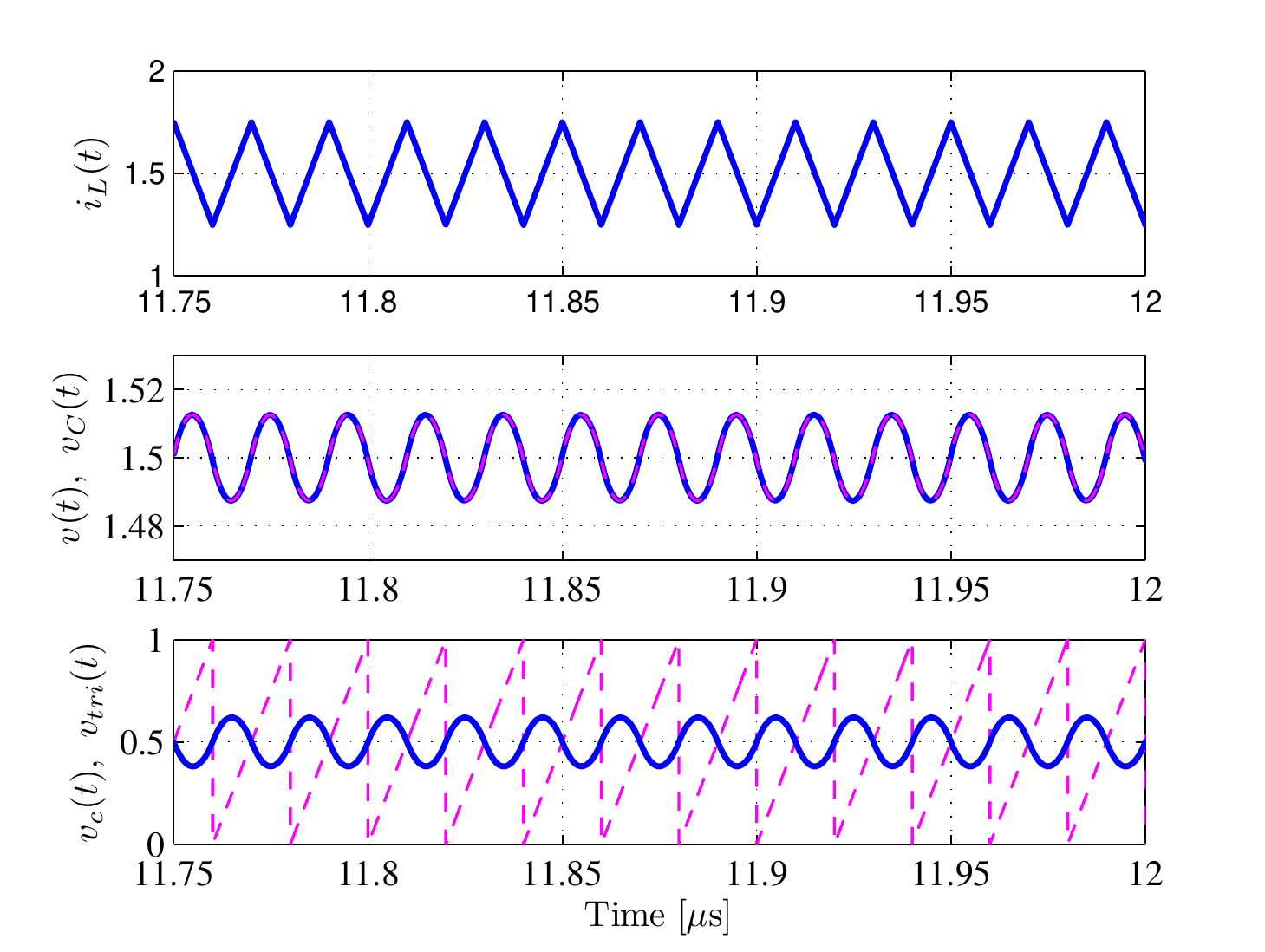}} \subfigure[$r_c~=~0.05\:\:\Omega$,  $r_\ell~=~1$ m$\Omega$,
$\alpha=1/2$, $k_v=10$]{\includegraphics[width=8cm]{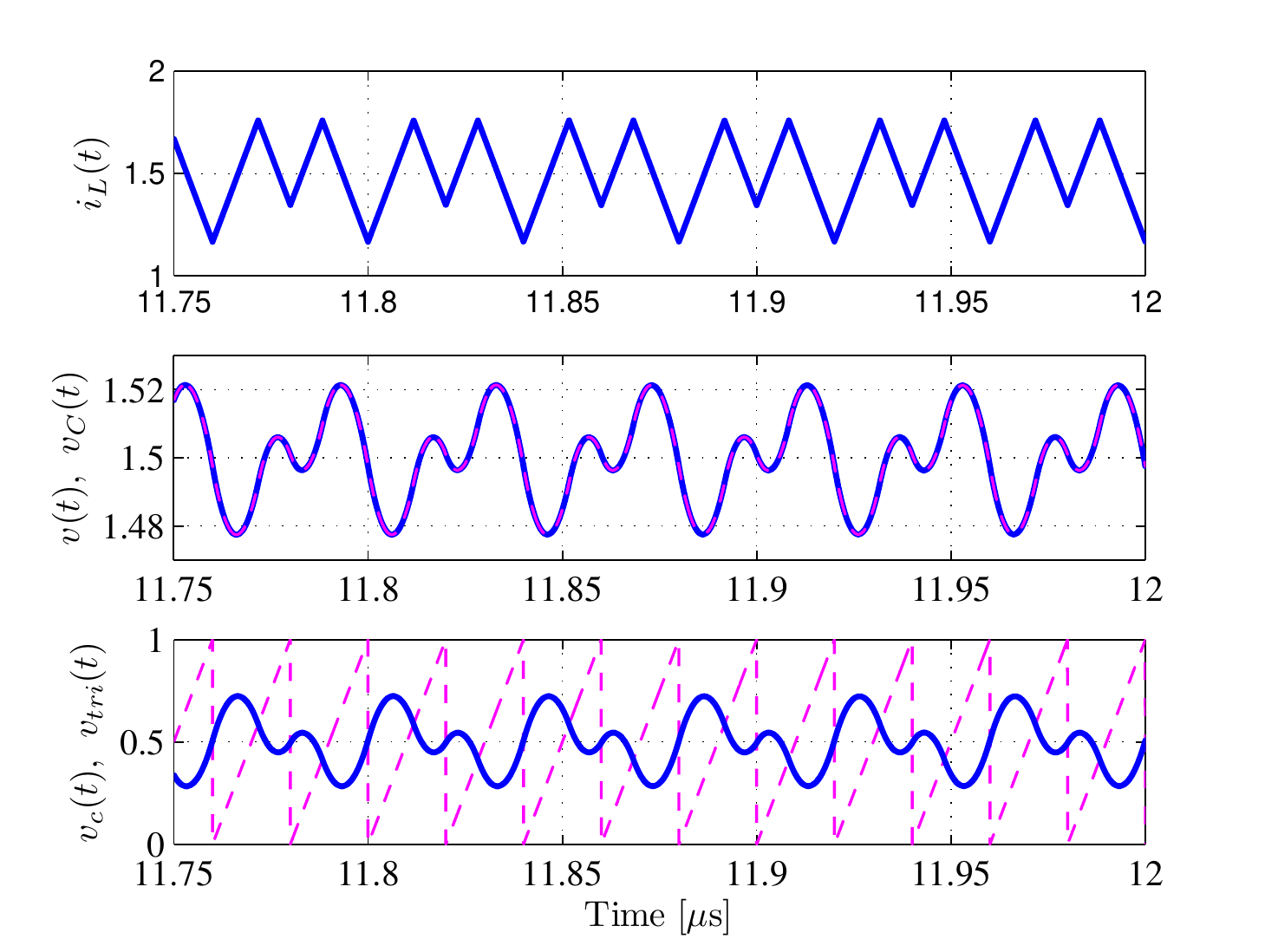}} \subfigure[$r_c~=~0$, $r_\ell=0$, $\alpha=1/2$,
$k_v=10$,]{\includegraphics[width=8cm]{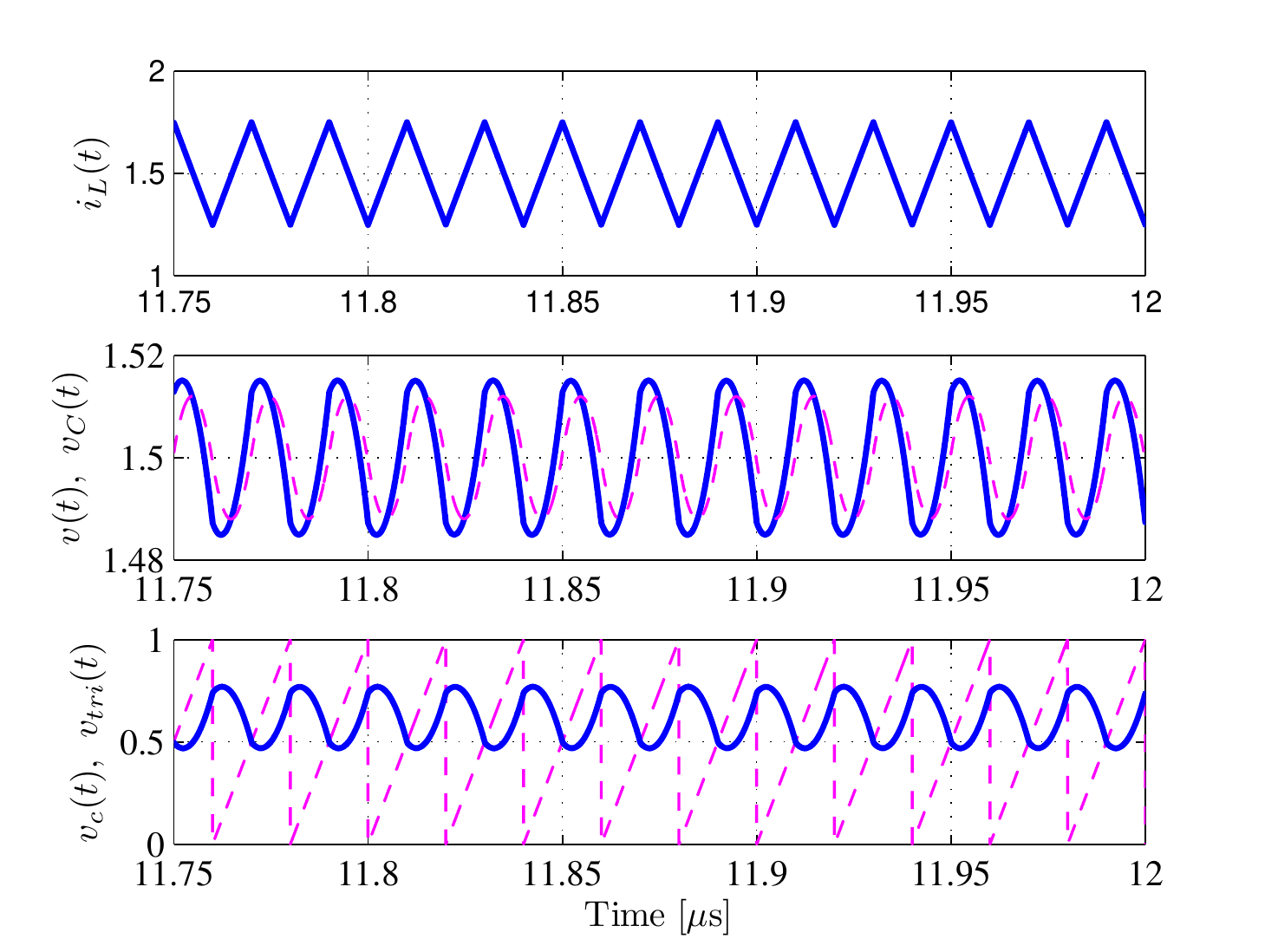}} \subfigure[$r_c~=~0.05\:\:\Omega$,  $r_\ell~=~1$ m$\Omega$, $\alpha=1/2$, $k_v=10.8$
]{\includegraphics[width=8cm]{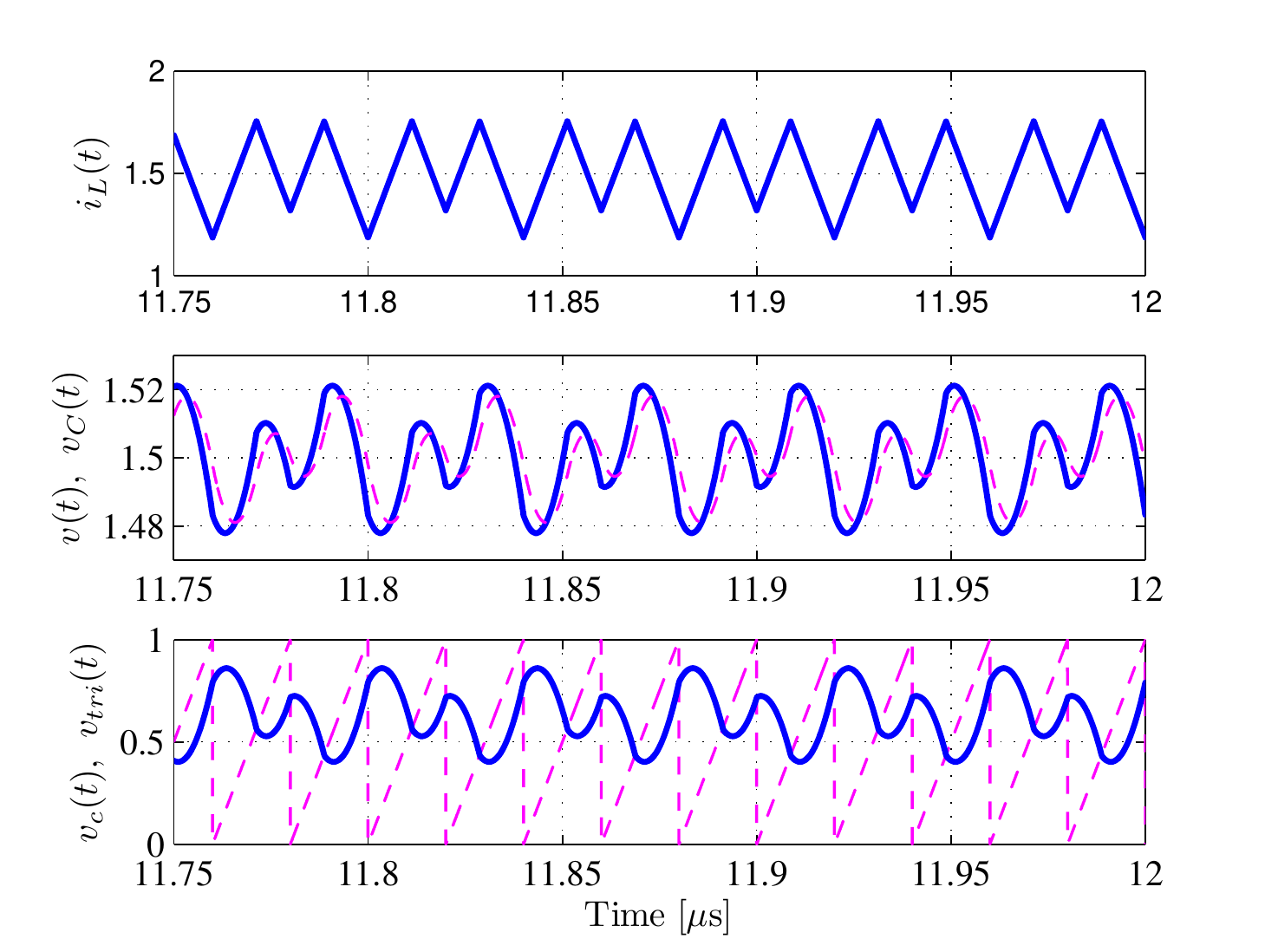}} \caption{Dynamic behavior of the buck converter of Example 2 just before and just after subharmonic oscillations
occurrence. (a) periodic behavior of the system with ideal components under a PI controller. (b) Subharmonic oscillation of the system under a PI controller with ideal
components. (c) Periodic behavior of the system under a Type II controller taking into account parasitic parameters. (d) Subharmonic oscillation of the system taking
into account parasitic parameters under a Type II controller.\label{fig:sim}}
\end{center}
\end{figure*}
\subsubsection{Taking into account parasitic elements}
Let us consider that $r_c=$ 0.05$\Omega$ and $r_\ell=$ 1 m$\Omega$. The system in this case is under a Type II controller. Its additional pole $\omega_{p1}$ is selected
at exactly $\omega_{z1}=1/r_cC$.  As before the zero $\omega_{z2}=\omega_0/2$ of the PI controller is below $\omega_0Q_0$. Figure \ref{fig:vg1vg2PI}-(b) shows the
stability curve in terms of the proportional gain $k_v$ and the duty cycle $D$ from \eqref{eq:vg2DbarDpi}. The critical value of the feedback gain for the previous set
of parameter values with $v_{ref}=~1.5$~V ($D=1/2$) is  $k_{v, {\rm cri}}\approx 10.7$.

Figure \ref{fig:sim}-(a)-(d) shows the time domain numerical simulation for the two previous cases for value of the feedback gain just before and just after subharmonic
oscillations occurrence. These numerical simulations are in perfect agreement with the theoretical derivations.
\end{example}

\section{Design-oriented prediction of subharmonic oscillations}
\subsection{Crossover frequency as a Figure of Merit}
The closed form results presented previously can be interpreted in terms of several Figures of Merit like for example the ripple amplitude in \cite{erla10}. The results
can be also interpreted by the crossover frequency and the associated phase margin which depends on the controller used. The crossover frequency $\omega_c$ is defined
as the frequency where the modulus of the loop gain crosses 0 dB ($|T_v(j\omega_c)| = 1$). For instance, for the buck converter under the proportional, PI and Type II
controllers, the crossover frequency can be approximated by
\begin{equation}
\omega_c\approx\omega_0\sqrt{\dfrac{k_v\kappa_\ell v_g}{V_M}} \label{eq:wcsub}
\end{equation}
Note that increasing the proportional gain $k_v$ or decreasing the modulator amplitude $V_M$ will raise the crossover frequency but this can also cause subharmonic
oscillations. From \eqref{eq:vg2DbarD} and \eqref{eq:wcsub}, the boundary conditions for this phenomenon, in terms of the crossover frequency can be written in the
following general form
\begin{equation}
\omega_c^2 \approx\omega_s^2\rho(D) \label{eq:vgwc}
\end{equation}
where $\rho(D)$ depends on the controller used. For the case of proportional, PI and Type II controllers and  if $\omega_0<<\omega_s$ and $\omega_0Q_0<<\omega_s$, from
\eqref{eq:vg2DbarD}, $\rho(D)$ can be approximated by
\begin{equation}
\rho(D)\approx \dfrac{1}{\pi^2(1-2D(1-D))}
\end{equation}
From \eqref{eq:wcsub}, the condition for the system to be free form subharmonic oscillations will be
\begin{equation}\omega_c<\omega_s\sqrt{\rho(D)}:=\omega_{c,{\rm
cri}}(D) \label{eq:wccri}
\end{equation}
In \cite{ridley}, the upper limit of the crossover frequency is discussed. For an ideal converter it was recommended, without demonstration, that an upper limit is
$\omega_s/5$. Eq. \eqref{eq:wccri} establish a theoretical upper limit for avoiding subharmonic oscillations in the buck converter.

For instance, in Example 1, taking  $v_{g}~=~23.5$~V and $R=22$~$\Omega$, the value of the duty cycle is $D~\approx~0.5$ and from \eqref{eq:wccri},
$\omega_c~=~6.95\:\:\text{krad/s}~<~\omega_{c,{\rm cri}}(D)~=~7.069\:\:\text{krad/s}$ and the system is stable. For $v_{g}~=~25.5$~V and $R=22$~$\Omega$, the value of
the duty cycle is $D~\approx~0.48$ and from \eqref{eq:wcsub}, $\omega_c~=~7.24\:\:\text{krad/s}~>~\omega_{c,{\rm cri}}(D)~=~7.059\:\:\text{krad/s}$ and therefore the
system exhibits subharmonic oscillation for this set of parameter values.

\subsection{Phase margin as a Figure of Merit}
Traditionally, the phase margin $\varphi_m$ is used, as a test on the loop gain $T_v(s)$, to determine the stability of the closed loop system
$T_{cl}(s)=T_v(s)/(1+T_v(s))$ in the averaging context. The phase margin $\varphi_m$ is determined from the phase of $T_v(s)$ at the crossover frequency $\omega_c$ , by
\begin{equation}
\varphi_m = \pi + \angle T_v(j\omega_c) \label{eq:phim}
\end{equation}
If $|T_v(s)|$ crosses 0 dB only once, then the closed loop system $T_v(s)/(1+T_v(s))$, the system $T_{cl}(s)$ will not present slow scale instability that can be
predicted by a simple averaged model. Traditionally a phase margin of at least $45^o$ is required to guarantee an acceptable system response \cite{erikson},
\cite{Hegarty}, \cite{Day}. Furthermore, to avoid subharmonic oscillations, the phase margin $\varphi_m$ must be greater than $\omega_{c,{\rm cri}}(D)$ in
\eqref{eq:wccri}.

Note that the phase margin will depend on the type of the controller used and the system parameters. From \eqref{eq:wccri} and \eqref{eq:phim}, the general expression
of the critical phase margin can be written in the following form
\begin{equation}
 \varphi_{m,{\rm cri}}(D) \approx \pi + \angle T_v(j\omega_{c,{\rm
cri}}(D))
\end{equation}
Subharmonic oscillations can be avoided if $\varphi_m<\varphi_{m,{\rm cri}}(D)$. For $\varphi_{m,{\rm cri}}(D)<45^o$, a well designed converter based on the loop gain
analysis of the averaged model will not exhibit subharmonic oscillations. For $\varphi_{m,{\rm cri}}(D)>45^o$, subharmonic oscillation may still occur even for a well
designed converter.

\section*{Conclusions} \label{sec:conclu}
Subharmonic oscillations in peak-current-mode-controlled converters are well documented  by different analytical methods. However, in voltage mode control, this
phenomenon has   been only characterized by using numerical simulations or mainly based on abstract mathematical analysis using a discrete time model or the Fillipov
method. Both approach do not allow to relate the results to concepts that are familiar to power electronics engineers. Explicit expressions for conditions of
subharmonic oscillations occurrence have been unavailable for many years.

It is widely known that for peak current mode control without ramp compensation, for example, the stability condition is $D<1/2$. This expression is shown to be a
special case of the general results presented in this paper. In general, it can be conjectured that the stability condition is in the form $f(D)<g(p)$, where $f$ is a
function of the duty cycle $D$ that can be approximated by a polynomial function and $g$ is a function of the vector of system parameters $p$. It was also shown that
the structure of the loop gain offers some additional insights on issues related to the stability of the system. Interestingly, the degree of the polynomial function
$f$ is extremely related to the relative degree of the total loop transfer function. For instance, for peak current mode control, the relative degree of the total loop
gain is 1 and the degree of $f$ is also 1. For voltage mode control, the relative degree of the loop gain depends on the compensator. For proportional controller,
single pole controller canceling the ESR zero of the power stage and for PI controller, the relative degree of the total loop is 2 and $f$ can also be approximated by
second degree polynomial function. The same can be conjectured for total loop transfer functions corresponding to other control schemes such as average current mode.

Based on the approach presented in this study, critical values of the system parameters can be located accurately.  Only in special limiting cases, an analytical
solution can be obtained. Analytical methods end at a certain point and have to be succeeded by numerics in the general case. However, hypergeometric-based functions
have some physical insights on how the system parameters affect the dynamics and how approximations should be done to predict subharmonic oscillations in an analytical
way in the case of a quality factor $Q_0>1/2$. Furthermore, by guaranteeing the stability for a high $Q_0$ case, the stability is also guaranteed for the low $Q_0$
case.

After explicitly obtaining the conditions for subharmonic oscillation, the results have been reformulated in terms of Figures of Merit widely used by the power
electronics community, such as the crossover frequency, and phase margin. The critical expressions for these Figures of Merit can be obtained in terms of converter
parameters and its controller. The ability to predict subharmonic instability using pure algebraic equations directly expressible in terms of such Figures of Merit is a
major advantage of the proposed approach.

Works are in progress to extend the approach to other switching converter topologies and other control schemes. The results will be reported in a further study.


\end{document}